\documentclass[a4paper,11pt]{article}
\pdfoutput=1 
\usepackage{jcappub}
\usepackage{amsmath}
\usepackage{bm}
\usepackage{graphicx}
\usepackage{enumitem}
\usepackage{geometry}
\usepackage{xspace}
\usepackage{pdflscape}
\usepackage{dsfont}

\usepackage{mathtools}

\usepackage{relsize}

\allowdisplaybreaks 

\makeatletter
\gdef\@fpheader{}
\g@addto@macro\bfseries{\boldmath}
\makeatother

\newcommand{\ie}{\textsl{i.e.~}}

\newcommand{\eg}{\textsl{e.g.~}}

\newcommand{\etc}{\textsl{etc.}}

\newcommand{\phiend}{\phi_{\text{end}}}
\newcommand{\phiuv}{\phi_{\text{uv}}}			

\newcommand{\dphiwell}{\Delta\phi_\mathrm{well}}

\newcommand{\N}{\mathcal{N}}

\let\oldsqrt\sqrt
\def\sqrt{\mathpalette\DHLhksqrt}
\def\DHLhksqrt#1#2{%
\setbox0=\hbox{$#1\oldsqrt{#2\,}$}\dimen0=\ht0
\advance\dimen0-0.2\ht0
\setbox2=\hbox{\vrule height\ht0 depth -\dimen0}%
{\box0\lower0.4pt\box2}}

\newcommand{\dd}{\mathrm{d}}
\newcommand{\ee}{e}

\newcommand{\sss}[1]{{\scriptscriptstyle{#1}}}
\newcommand{\boldmathsymbol}[1]{{\ensuremath{\boldsymbol{#1}}}}

\newcommand{\uPl}{\mathrm{Pl}}
\newcommand{\uin}{\mathrm{in}}

\newcommand{\uend}{\mathrm{end}}

\newcommand{\ucl}{\mathrm{cl}}

\newcommand{\uc}{\mathrm{c}}

\newcommand{\usssPl}{\sss{\uPl}}

\newcommand{\uNL}{\mathrm{NL}}

\newcommand{\calP}{\mathcal{P}}

\newcommand{\Mp}{M_\usssPl}

\newcommand{\fnl}{f_\uNL}
\newcommand{\gnl}{g_\uNL}

\newcommand{\efolds}{$e$-folds}

\newcommand{\beq}{\begin{equation}}
\newcommand{\eeq}{\end{equation}}
\newcommand{\bea}{\begin{equation}\begin{aligned}}
\newcommand{\eea}{\end{aligned}\end{equation}}

\newlength{\wsingfig}
\newlength{\wdblefig}
\newlength{\wquadfig}
\newlength{\wtriplefig}
\newcommand{\Eq}[1]{Eq.~(\ref{#1})}
\newcommand{\Eqs}[1]{Eqs.~(\ref{#1})}
\newcommand{\Fig}[1]{Fig.~{\ref{#1}}}

\newcommand{\Ref}[1]{Ref.~{\cite{#1}}}
\newcommand{\Refs}[1]{Refs.~{\cite{#1}}}
\newcommand{\Sec}[1]{Sec.~\ref{#1}}
\newcommand{\Secs}[1]{Secs.~\ref{#1}}
\newcommand{\App}[1]{Appendix~\ref{#1}}

\newcommand{\lp}{\left(}
\newcommand{\rp}{\right)}
\newcommand{\lb}{\left[}
\newcommand{\rb}{\right]}

\newcommand{\deflen}[2]{%
    \expandafter\newlength\csname #1\endcsname
    \expandafter\setlength\csname #1\endcsname{#2}%
}

\definecolor{orange}{rgb}{1,0.5,0}

\newcommand{\zetacg}{\zeta_{\mathrm{cg}}}
\newcommand{\zetac}{\zeta_{\mathrm{c}}}

\subheader{}

\title{The exponential tail of inflationary fluctuations: consequences for primordial black holes}

\author[a]{Jose Mar\'ia Ezquiaga,}
\author[b]{ Juan Garc\'ia-Bellido,}
\author[c]{Vincent Vennin}

\affiliation[a]{NASA Einstein Fellow, Kavli Institute for Cosmological Physics and Enrico Fermi Institute, The University of Chicago, Chicago, IL 60637, USA}

\affiliation[b]{Instituto de F\'isica Te\'orica UAM-CSIC, Universidad Aut\'onoma de Madrid, Cantoblanco,
Madrid, 28049 Spain}

\affiliation[c]{Laboratoire Astroparticule et Cosmologie, Universit\'e
  Denis Diderot Paris 7, 10 rue Alice Domon et L\'eonie Duquet, 
75013 Paris, France}

\emailAdd{ezquiaga@uchicago.edu}
\emailAdd{juan.garciabellido@uam.es}
\emailAdd{vincent.vennin@apc.univ-paris7.fr}

\date{today}

\begin{document}
\sloppy

\abstract{
The curvature perturbations produced during an early era of inflation are known to have quasi-Gaussian distribution functions close to their maximum, where they are well constrained by measurements of the cosmic microwave background anisotropies and of the large-scale structures. In contrast, the tails of these distributions are poorly known, although this part is the relevant one for rare, extreme objects such as primordial black holes. We show that these tails are highly non-Gaussian, and cannot be described with standard non-Gaussian expansions, that are designed to approximate the distributions close to their maximum only. Using the stochastic-$\delta N$ formalism, we develop a generic framework to compute the tails, which are found to have an exponential, rather than Gaussian, decay. These exponential tails are inevitable, and do not require any non-minimal feature as they simply result from the quantum diffusion of the inflaton field along its potential. We apply our formalism to a few relevant single-field, slow-roll inflationary potentials, where our analytical treatment is confirmed by comparison with numerical results. We discuss the implications for the expected abundance of primordial black holes in these models, and highlight that it can differ from standard results by several orders of magnitude. In particular, we find that potentials with an inflection point overproduce primordial black holes, unless slow roll is violated.}

\maketitle
%
\section{Introduction}
The primordial inhomogeneities produced out of vacuum quantum fluctuations during an early phase of inflation seed all cosmological structures in our universe~\cite{Starobinsky:1979ty, Starobinsky:1980te, Guth:1980zm, Linde:1981mu, Albrecht:1982wi, Linde:1983gd, Mukhanov:1981xt, Mukhanov:1982nu, Starobinsky:1982ee, Guth:1982ec, Hawking:1982cz, Bardeen:1983qw}. They can be measured in the temperature and polarisation anisotropies of the cosmic microwave background (CMB)~\cite{Akrami:2018vks} at gigaparsec scales, and in probes of the large-scale structure of our universe at megaparsec scales~\cite{Amendola:2012ys, Amendola:2016saw}. They are constrained to be quasi-scale invariant, quasi adiabatic and quasi-Gaussian~\cite{Aghanim:2018eyx}. Current constraints on the amount of non-Gaussianities come from upper bounds on the amplitude of the bispectrum and the trispectrum~\cite{Akrami:2019izv}, and therefore capture deviations from Gaussian statistics close to the maximum of the distribution functions, the width of which is given by measurements of the power spectrum.

At smaller scales, large quantum fluctuations could also give rise to rare and more extreme objects such as ultra compact mini-halos, or primordial black-holes (PBH)~\cite{Carr:1974nx,GarciaBellido:1996qt}.
Even if sparsely distributed, PBHs are of particular cosmological relevance as they could provide seeds for supermassive black-holes in galactic nuclei~\cite{Bean:2002kx, Clesse:2015wea, Carr:2018rid}, and comprise a large fraction of the dark matter~\cite{Clesse:2017bsw, Carr:2019kxo}. There has been renewed interest in PBHs since the LIGO/VIRGO collaboration reported the first detection of gravitational waves associated to black-hole mergers in 2015~\cite{Abbott:2016blz}, as they may indeed explain the existence of progenitors for these events~\cite{Bird:2016dcv, Clesse:2016vqa, Sasaki:2016jop}. PBHs may also play a role in the generation of large-scale structures~\cite{Meszaros:1975ef, Carr:2018rid} and solve a number of problems currently encountered in astrophysics and cosmology (see \eg \Ref{Clesse:2017bsw} for hints in favour of the existence of PBHs).

Constraints on the abundance of PBHs have been placed in various mass ranges (see \eg \Refs{Carr:2009jm, Carr:2017jsz} for reviews), since PBHs leave several imprints throughout the history of the universe: by injecting energy into their surrounding environment, by providing a source of gravitational lensing, by sourcing various dynamical effects, and by emitting gravitational waves~\cite{Carr:2016drx, Garcia-Bellido:2017fdg, Sasaki:2018dmp}. This leaves two mass windows open for PBHs to constitute an appreciable fraction, and possibly all, of dark matter, around $M\sim 10^{-12}M_\odot$ and $M\sim 10-100 M_\odot$, where $M_\odot$ denotes the mass of the sun. 

PBHs thus constitute a new observational window. Indeed, the CMB only gives access to a limited range of scales, and the time frame over which these scales are generated during inflation is therefore limited as well, and cannot encompass more than $\sim$ 7 $e$-folds (over the $\sim 60$ $e$-folds elapsed between the generation of these scales and the end of inflation). This means that the constraints on the inflationary potential that the CMB can place are restricted to a small field range. By giving access to much smaller scales, PBHs could thus allow us to probe parts of the inflationary potential that would remain hidden to us otherwise.

A PBH is expected to form when a large density perturbation re-enters the Hubble radius after inflation, and collapses into a black hole if it exceeds a certain threshold. More precisely, denoting by $\zetacg$ the curvature perturbation coarse-grained above a certain scale $\bar{k}$ and the end of inflation $k_\text{end}=a_\text{end}H_\text{end}$, $\zetacg(\bm{x}) = \left(2\pi\right)^{-3/2}\int_{k_\text{end}>k>\bar{k}}\dd {\bm{k}}\zeta_{\bm{k}} e^{i\bm{k}\cdot\bm{x}}$ (see \App{sec:zetacg}), and by $P(\zetacg)$ the probability density function (PDF) of $\zetacg$, the fraction of the universe made of PBHs with mass $M(\bar{k})$, at their formation time, is given by~\cite{1975ApJ...201....1C}
\bea
\label{eq:beta:pdf}
\beta_{\mathrm{f}}(M)=\int_{\zeta_{\rm c}}^{\infty}P(\zetacg)\dd \zetacg\, .
\eea
Here, one can relate the mass $M(\bar{k})$ with the mass contained within the Hubble radius $H^{-1}=a/\dot{a}$ at the time of formation, \ie when $\bar{k}=aH$, where $a$ is the scale factor and a dot denotes derivation with respect to cosmic time. In this expression, $\zetac$ is the threshold value for forming a black hole and is typically of order one. Since the typical values of $\zetacg$ are much smaller than one (perturbations remaining in the linear regime on super-Hubble scales), \Eq{eq:beta:pdf} involves an integral over the high-curvature tail of the PDF. This has two important implications. First, the observational constraints mentioned above, limiting the deviations from Gaussian statistics close to the maximum of the PDFs at large scales, are not directly relevant for describing the tails of distributions at much smaller scales. Second, all techniques developed in the literature to compute the PDF of $\zetac$ from inflation are designed to provide approximations modelling the neighbourhood of the maximum of the PDF, not its tail. This is for instance the case for the expansion in terms of the non-linearity parameters $f{{}_\mathrm{NL}}$ and $g{{}_\mathrm{NL}}$~\cite{Gangui:1993tt}. 

Non-Gaussianities may therefore play a crucial role in determining the abundance of PBHs~\cite{Byrnes:2012yx, Young:2013oia, Young:2015cyn, Garcia-Bellido:2016dkw, Garcia-Bellido:2017aan, Ezquiaga:2018gbw, Franciolini:2018vbk,Cai:2018dig, Passaglia:2018ixg, Young:2019yug, DeLuca:2019qsy, Panagopoulos:2019ail, Yoo:2019pma,Carr:2019hud} and, in this work, we provide a generic, non-perturbative framework to derive the tails of the PDFs of curvature perturbations. This makes use of the stochastic-$\delta N$ formalism~\cite{Enqvist:2008kt, Fujita:2013cna, Fujita:2014tja, Vennin:2015hra, Kawasaki:2015ppx, Assadullahi:2016gkk, Vennin:2016wnk, Pinol:2018euk}, in which the curvature perturbations are identified with fluctuations in the local duration of inflation, that varies under the effects of quantum diffusion of the inflaton field, described as stochastic noise in the stochastic inflation formalism~\cite{Starobinsky:1982ee, Starobinsky:1986fx, Linde:1993xx, Starobinsky:1994bd}. We find that this stochastic noise gives rise to exponentially decaying tails, that are highly enhanced compared to Gaussian tails. This property is universal, and arises as soon as quantum fluctuations of the background inflaton are incorporated into the analysis; it does not involve any non-minimal feature and occurs in all scenarios. The reason why quantum diffusion plays a crucial role in determining the abundance of PBHs is because PBHs form in regimes where large curvature perturbations, exceeding the threshold, can be produced. This requires a very flat inflationary potential, where the dynamics of the inflaton is no longer dominated by the classical drift (induced by the potential gradient), but rather by the stochastic noise. 

The paper is organised as follows. In \Sec{sec:tail_curvature}, we first review the stochastic-$\delta N$ formalism. For single-field slow-roll potentials, an exact analytical solution for the PDF of curvature perturbations can be found only in the case of an exactly flat potential~\cite{Pattison:2017mbe}. This is why we develop a tail expansion that can be used in arbitrary potentials, and that relies on computing the poles of the characteristic function of the PDF. In \Sec{sec:applications}, we apply our computational program to a few potentials: exactly flat potentials (for which the full solution of \Ref{Pattison:2017mbe} is recovered, although derived in a different way) in \Sec{sec:flat_potential}, potentials with a constant slope in \Sec{sec:linear_potential}, potentials with a cubic flat inflection point in \Sec{sec:inflection_potential}, and potentials with a linearly-tilted cubic inflection point in \Sec{sec:inflection_linear_potential}. The implications for the expected amount of PBHs are discussed in \Sec{sec:pbh}, where it is shown that quantum diffusion, and non-Gaussian tails, can change the abundance of PBHs by several orders of magnitude compared to standard results, and can be such that substantial abundances can be reached even in the absence of slow-roll violations. Finally, in \Sec{sec:discussion}, we present some concluding remarks.
\section{The tail of the curvature perturbations distribution}
\label{sec:tail_curvature}
In this section, we explain how the tail of the PDF of curvature perturbations can be computed in the stochastic-$\delta N$ formalism. We consider the case where inflation is driven by one or several scalar fields $\phi_i$, with $i=1\cdots n$, and the action of the system if given by
\bea
	S=\displaystyle\int \dd^4x\sqrt{-g}\left[\frac{\Mp^2}{2}R-\frac{1}{2}\delta^{ij}g^{\mu\nu}\partial_\mu\phi_i\partial_\nu\phi_j-V(\phi_i)\right] , 
	\label{eq:action}
\eea
where $\Mp$ is the reduced Planck mass. If time is labeled by the number of \efolds~$N\equiv \ln a$, for each field $\phi_i$, one can define a conjugate momentum $\pi_i=\dd\phi_i/\dd N$, and phase space is parametrised with the field vector $\boldmathsymbol{\Phi}=(\phi_1,\pi_1,\cdots,\phi_n,\pi_n)$.
\subsection{The stochastic-$\delta N$ formalism}
\label{sec:Stochastic:DeltaN}
In the $\delta N$ formalism~\cite{Starobinsky:1982ee, Starobinsky:1986fxa, Sasaki:1995aw, Sasaki:1998ug, Lyth:2004gb, Lyth:2005fi}, the curvature perturbations on large scales are nothing but the fluctuations in the number of \efolds~of expansion during inflation for a family of homogeneous universes. Indeed, in a gauge where fixed time slices have uniform energy density and fixed spatial worldlines are comoving (in the super-Hubble limit, this gauge reduces to the synchronous gauge supplemented by some additional conditions that fix it uniquely), the perturbed flat Friedmann-Lema\^{i}tre-Robertson-Walker (FLRW) metric reads ~\cite{Starobinsky:1982ee, Creminelli:2004yq,Salopek:1990jq} $\dd s^2 = -\dd t^2 + a^2(t)\ee^{2\zeta(t, \bm{x})}\delta_{ij}\dd x^{i}\dd x^{j}$, where only scalar fluctuations have been included. A local scale factor can then be introduced, $\tilde{a}(t, \bm{x}) = a(t)\ee^{\zeta(t, \bm{x})}$, which allows us to relate the amount of expansion from an initial flat space-time slice at time $t_\uin$ to a final space-time slice of uniform energy density as $N(t, \bm{x}) = \ln{\left[ \tilde{a}(t, \bm{x})/a(t_{\mathrm{in}}) \right]}$. This gives rise to
\bea
\label{eq:zeta:deltaN}
\zeta(t, \bm{x}) = N(t, \bm{x}) - \bar{N}(t) \equiv \delta N \, ,
\eea
where $\bar{N}(t) \equiv \ln{\left[ {a(t)}/{a(t_{\mathrm{in}})} \right]}$ is the unperturbed expansion. Moreover, on super-Hubble scales, gradients can be neglected, and each spatial point evolves independently and follows the evolution of an unperturbed universe.  This is known as the ``quasi-isotropic''~\cite{Lifshitz:1960, Starobinsky:1982mr, Comer:1994np, Khalatnikov:2002kn} or ``separate universe'' approach~\cite{Wands:2000dp, Lyth:2003im, Lyth:2004gb}, the validity of which has recently been shown to extend beyond slow roll~\cite{Pattison:2019hef}. As a consequence, $N(t, \bm{x})$ is the amount of expansion in unperturbed, homogeneous universes, and $\zeta$ can be calculated from the knowledge of the evolution of a family of such universes. 

The calculation of the PDF of curvature perturbations therefore boils down to the calculation of the PDF of local durations of inflation. These durations vary under quantum fluctuations in the fields that drive inflation, which can be described in the framework of stochastic inflation~\cite{Starobinsky:1982ee, Starobinsky:1986fx, Linde:1993xx, Starobinsky:1994bd}. This formalism is an effective theory for the long wavelength part of the fields, which are coarse-grained below a fixed physical scale $k_\sigma=\sigma a H$, 
\bea
\hat{\boldmathsymbol{\Phi}}_{\mathrm{cg}}(\bm{x}) = \frac{1}{\left(2\pi\right)^{3/2}}\int_{k<k_\sigma} \dd^3 k \hat{\boldmathsymbol{\Phi}}_k \ee^{-i k \bm{x}} ,
\eea
where $\sigma\ll 1$ is a fixed parameter setting the scale at which quantum fluctuations backreact onto the local FLRW geometry. The dynamics of the quantum operators $\hat{\boldmathsymbol{\Phi}}_{\mathrm{cg}}$ can be tracked by means of stochastic Langevin equations,
\bea
\label{eq:Langevin}
\frac{\dd {\boldmathsymbol{\Phi}}_{\mathrm{cg}}}{\dd N} = \boldmathsymbol{F}_{\mathrm{cl}}\left(\boldmathsymbol{\Phi}_{\mathrm{cg}}\right)+\boldmathsymbol{\xi},
\eea
where $\boldmathsymbol{F}_{\mathrm{cl}}\left(\boldmathsymbol{\Phi}\right)$ encodes the classical equations of motion [and is given by the commutator between $\boldmathsymbol{\Phi}$ and the Hamiltonian associated to \Eq{eq:action}], and $\boldmathsymbol{\xi}$ is a white Gaussian noise with vanishing mean, and variance given by
\bea
\left\langle \xi_i\left(\bm{x}_i,N_i\right) \xi_j\left(\bm{x}_j,N_j\right) \right\rangle = \frac{\dd\ln k_\sigma}{\dd N} \calP_{\Phi_i,\Phi_j}\left[k_\sigma(N_i),N_i\right]\delta(N_i-N_j) .
\eea
In this expression, $\calP_{\Phi_i,\Phi_j}[k_\sigma(N_i),N_i]\delta(N_i-N_j)$ is the cross power spectrum between the field variables $\Phi_i$ and $\Phi_j$, evaluated at the scale $k_\sigma(N_i)$, and at time $N_i$. Note that since the time coordinate (here $N$) has not been perturbed in the Langevin equation~(\ref{eq:Langevin}), we are implicitly working in the gauge where time is unperturbed, \ie the uniform $N$-gauge in the present case. The power spectra need therefore to be computed in that gauge~\cite{Pattison:2019hef}. 

The Langevin equation~(\ref{eq:Langevin}) thus describes a family of background histories, each of them realising a different number of inflationary \efolds. The stochastic-$\delta N$ formalism~\cite{Enqvist:2008kt, Fujita:2013cna, Fujita:2014tja, Vennin:2015hra, Kawasaki:2015ppx, Assadullahi:2016gkk, Vennin:2016wnk, Pinol:2018euk} then consists in computing the PDF of this number of \efolds, from a certain initial configuration in field space until inflation terminates, \ie until the surface $\mathcal{C}_\uend=\lbrace \boldmathsymbol{\Phi} \vert \epsilon_{1}(\boldmathsymbol{\Phi})=1 \rbrace$ is crossed out, where $\epsilon_1=-\dot{H}/H^2$. This is done by deriving the Fokker-Planck equation associated to the Langevin equation~(\ref{eq:Langevin}), that drives the probability to find the system $\boldmathsymbol{\Phi}_{\mathrm{cg}}$ at position $\boldmathsymbol{\Phi}$ in field space at time $N$, knowing that it was at position $\boldmathsymbol{\Phi}_\uin$ at a previous time $N_\uin$,
\bea
\frac{\dd }{\dd N}P\left(\boldmathsymbol{\Phi},N\vert \boldmathsymbol{\Phi}_\uin,N_\uin\right) = \mathcal{L}_{\mathrm{FP}}\left(\boldmathsymbol{\Phi}\right)\cdot P\left(\boldmathsymbol{\Phi},N\vert \boldmathsymbol{\Phi}_\uin,N_\uin\right) .
\eea
Here, $ \mathcal{L}_{\mathrm{FP}}\left(\boldmathsymbol{\Phi}\right)$ is the Fokker-Planck operator, which is a differential operator of second order in phase space (\ie it contains first and second-order derivatives with respect to the field coordinates $\Phi_i$). The PDF of the number of \efolds~realised from $\boldmathsymbol{\Phi}$ until the end of inflation, $\N$, can then be shown to obey the adjoint Fokker-Planck equation~\cite{Vennin:2015hra, Pattison:2017mbe}
\bea
\label{eq:adjoint:FP}
\frac{\dd }{\dd \mathcal{N}} P_{\boldmathsymbol{\Phi}}\left(\N\right) = \mathcal{L}_{\mathrm{FP}}^\dagger \left(\boldmathsymbol{\Phi}\right)\cdot P_{\boldmathsymbol{\Phi}}\left(\N\right) ,
\eea
where $\mathcal{L}_{\mathrm{FP}}^\dagger(\boldmathsymbol{\Phi})$ is the adjoint Fokker-Planck operator, related to the Fokker-Planck operators via $\langle F_1, \mathcal{L}(F_2)\rangle \equiv$
$\int \dd \boldmathsymbol{\Phi} F_1(\boldmathsymbol{\Phi}) \mathcal{L}_{\mathrm{FP}}(\boldmathsymbol{\Phi})\cdot F_2(\boldmathsymbol{\Phi}) = \int \dd \boldmathsymbol{\Phi} F_2(\boldmathsymbol{\Phi}) \mathcal{L}_{\mathrm{FP}}^\dagger(\boldmathsymbol{\Phi})\cdot F_1(\boldmathsymbol{\Phi}) =\langle \mathcal{L}_{\mathrm{FP}}^\dagger(F_1),F2\rangle$. For instance, in the case where the dynamics of all fields proceed in the slow-roll regime, phase space can be parametrised by the field values $\phi_i$ only and the momenta $\pi_i$ can be dropped from the vector $\boldmathsymbol{\Phi}$, and one has~\cite{Vennin:2015hra, Assadullahi:2016gkk}
\bea
\label{eq:FP:SR}
\frac{1}{\Mp^2}\mathcal{L}_{\mathrm{FP}}^\dagger &= -\sum_i \frac{v_{\phi_i}}{v}\frac{\partial}{\partial\phi_i} + v \sum_i \frac{\partial^2}{\partial\phi_i^2}\,, \\
\eea
where
\bea
\label{eq:reduced:potential}
v=\frac{V}{24\pi^2\Mp^4}
\eea
denotes the reduced potential. In this section, we remain generic and do not assume slow roll, while in \Sec{sec:applications}, the case of single-field slow-roll models will be considered, hence \Eq{eq:FP:SR} will be used. Let us also note that \Eq{eq:adjoint:FP} needs to be solved with some boundary conditions. As already mentioned, inflation terminates on the surface $\mathcal{C}_\uend$, so a first condition is that $P_{\boldmathsymbol{\Phi}}\left(\N\right) = \delta(\N)$ when $\boldmathsymbol{\Phi} \in \mathcal{C}_\uend$. For completeness, we also introduce reflective boundary conditions high in the potential~\cite{Assadullahi:2016gkk, Vennin:2016wnk} (the interpretation of which will be made clearer below), across some surface $\mathcal{C}_{\mathrm{uv}}$, such that $[\bm{u}(\boldmathsymbol{\Phi})\cdot\bm{\nabla}]P_{\boldmathsymbol{\Phi}}\left(\N\right)=0$ when $\boldmathsymbol{\Phi} \in \mathcal{C}_{\mathrm{uv}}$, where $\bm{u}$ is the field-space vector orthogonal to $\mathcal{C}_{\mathrm{uv}}$. From the knowledge of the PDF of $\N$, the PDF of the coarse-grained curvature perturbation,
\bea \label{eq:cg-deltaN}
\zetacg\left(\bm{x}\right)=
\delta N_{\mathrm{cg}}\left(\bm{x}\right) = 
\mathcal{N}\left(\bm{x}\right)-\left\langle \mathcal{N} \right\rangle ,
\eea
can finally be obtained.

Let us stress that although $P(\boldmathsymbol{\Phi},N\vert \boldmathsymbol{\Phi}_\uin,N_\uin) $ and $P_{\boldmathsymbol{\Phi}}(\N)$ obey very similar equations, they are conceptually very different objects. In what follows, we will show that for large values of $\N$, the latter admits an expansion of the form
\bea
\label{eq:tail_expansion}
P_{\boldmathsymbol{\Phi}}(\N)=\sum_{n}a_n(\boldmathsymbol{\Phi})e^{-\Lambda_n\,\N}\,,
\eea
where the functions $a_n(\boldmathsymbol{\Phi})$ determine the amplitude of the tail, and the coefficients $\Lambda_n$, which we will show do \emph{not} depend on $\boldmathsymbol{\Phi}$, set the exponential decay rates. Notice that in the classical, Gaussian picture, the PDF is given by
\bea
\label{eq:tail_expansion:classical}
\left. P_{\boldmathsymbol{\Phi}}(\N)\right\vert_{\ucl} 
\underset{\mathcal{N}\gg 1}{\propto}
 \exp\left[-\frac{1}{2}\frac{\mathcal{N}^2}{\int_{\bar{k}}^{k_\uend} \calP_{\zeta,\mathrm{cl}}(k)\dd\ln k}\right]\,,
\eea
where $\calP_{\zeta,\mathrm{cl}}$ is the classical value of the power spectrum [in single-field slow-roll inflation, it is given by $\calP_{\zeta,\mathrm{cl}}=2v^3/(\Mp^2v_\phi^2)$], and $\bar{k}$ and  $k_\uend$ are the scales that cross out the Hubble radius when the system is at location $\boldmathsymbol{\Phi}$, and at the end of inflation, respectively. If non-Gaussianities are perturbatively introduced by the means of the usual non-linearity parameters $f{{}_\mathrm{NL}}$, $g{{}_\mathrm{NL}}$ \etc , \Eq{eq:tail_expansion:classical} is modified with polynomial corrections in $\N$~\cite{Vennin:2015hra, Vennin:2016wnk}, which cannot capture the exponential decay of \Eq{eq:tail_expansion}. 

We now present two complementary techniques to compute $a_n(\boldmathsymbol{\Phi})$ and $\Lambda_n$, before applying them to concrete examples in \Sec{sec:applications}.
\subsection{Poles of the characteristic function }
\label{sec:pole:chi}
In order to analyse the solution of \Eq{eq:adjoint:FP} in the large-$\N$ limit, it is convenient~\cite{Pattison:2017mbe} to introduce the characteristic function
\bea
\label{eq:characteristicFunction:def}
\chi_\N\left(t,\boldmathsymbol{\Phi}\right) \equiv \left\langle \ee^{it \N(\boldmathsymbol{\Phi})} \right\rangle
=\int_{-\infty}^{\infty}e^{it\N}
P_{\boldmathsymbol{\Phi}}\left(\N\right)
 \dd \N ,
\eea 
where $t$ is a dummy parameter. From the second equality, one can see that the characteristic function is nothing but the Fourier transform of the PDF, hence the PDF can be obtained by inverse Fourier transforming the characteristic function,
\bea
\label{eq:pdf:chi}
P_{\boldmathsymbol{\Phi}}\left(\N\right)=\frac{1}{2\pi}\int_{-\infty}^{\infty}e^{-it\N}\chi_\N\left(t,\boldmathsymbol{\Phi}\right)\dd t\, .
\eea
By plugging \Eq{eq:pdf:chi} into \Eq{eq:adjoint:FP}, one obtains
\bea
\label{eq:diff:chi}
\left[\mathcal{L}_{\mathrm{FP}}^\dagger \left(\boldmathsymbol{\Phi}\right) +it\right]\chi_\N\left(t,\boldmathsymbol{\Phi}\right)= 0\,
\eea
with boundary conditions $\chi_\N(t,\boldmathsymbol{\Phi})=1$ when $\boldmathsymbol{\Phi} \in \mathcal{C}_\uend$, and $[\bm{u}(\boldmathsymbol{\Phi})\cdot\bm{\nabla}]\chi_\N(t,\boldmathsymbol{\Phi})=0$ when $\boldmathsymbol{\Phi} \in \mathcal{C}_{\mathrm{uv}}$.
\begin{figure}[t]
\centering 
\includegraphics[width=.49\textwidth]{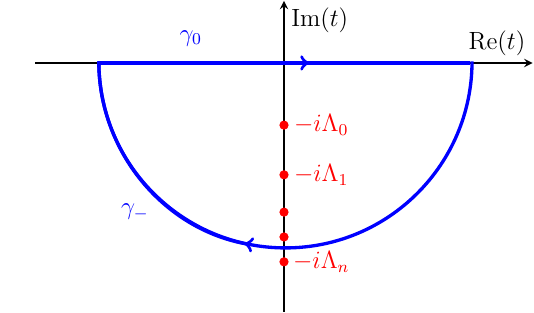}
 \caption{Schematic representation of the pole structure of the characteristic function. In order to compute the PDF, $P_{\boldmathsymbol{\Phi}}(\N)$, from the characteristic function, making use of the residue theorem, the real axis integral of \Eq{eq:pdf:chi} can be obtained from the integral over the contour $\gamma_0\cup\gamma_{-}$ in the complex plane.}
 \label{fig:contour_poles}
\end{figure}

The idea is to perform the integral of \Eq{eq:pdf:chi} by means of the residue theorem.\footnote{If $f(z)$ is a regular function in the complex plane, and $\gamma$ a close contour that circles in a certain point $z_p$ (the winding number of $\gamma$ around $z_p$ is one), one has
\bea
\oint_{\gamma}\frac{f(z)\,\dd z}{(z-z_p)^{n+1}}=\frac{2\pi i}{n!}\lb\frac{\dd ^n}{\dd z^n}f(z)\rb_{z=z_p}\, .
\eea
}
This is done by expanding the characteristic function according to
\bea
\label{eq:chi:pole:expansion}
\chi_\N\left(t,\boldmathsymbol{\Phi}\right)=\sum_{n}\frac{a_n(\boldmathsymbol{\Phi})}{\Lambda_n-it}+g(t,\boldmathsymbol{\Phi})\,,
\eea
where $g(t,\boldmathsymbol{\Phi})$ is a regular function of $t$, and the $\Lambda_n$ are positive numbers that do not depend on $\boldmathsymbol{\Phi}$. The form of this expansion can be justified as follows.\footnote{We thank Nahid Ahmadi, Zahra Ahmadi, Niloufar Feyz, Mahdiyar Noorbala and Borna Salehian for pointing out that the argument given for the validity of the expansion~\eqref{eq:chi:pole:expansion} in a previous version of the paper was incomplete, and for engaging in an insightful discussion that led to the considerations presented in this paragraph.\label{footnote:ack}}
First, since the PDF $P_{\boldmathsymbol{\Phi}}(\mathcal{N})$ vanishes for negative $\mathcal{N}$, according to Titchmarsh's theorem~\cite{Titchmarsh:1937kpd}, its Fourier transform, \ie the characteristic function, is a holomorphic function in the upper half-plane. Therefore, the poles of the characteristic functions all have a negative imaginary part. Second, since the PDF $P_{\boldmathsymbol{\Phi}}(\mathcal{N})$ is a real fuction of $\mathcal{N}$, from \Eq{eq:characteristicFunction:def} its Fourier transform must satisfy $\chi_\mathcal{N}^*(t,\boldmathsymbol{\Phi})=\chi_\mathcal{N}(-t^*,\boldmathsymbol{\Phi})$. If a pole of the characteristic function existed outside the negative imaginary axis, this would give rise to a pole on the upper half-plane, which is forbidden, hence all poles lie on the negative imaginary axis at locations $t=-i\Lambda_n$. In what follows, the $\Lambda_n$ are ordered such that $0<\Lambda_0<\Lambda_1<\Lambda_2<\cdots<\Lambda_n$. A priori, there is no reason why these poles should be all simple poles, but in \Sec{sec:eigenvalues} we will show that this is indeed the case. Third, \Eq{eq:diff:chi} is a second-order linear differential equation, which is further linear in $t$, so there exist independent solutions that are regular in $t$, and the poles in $t$ only appear when enforcing the boundary conditions. This explains why the $\Lambda_n$ do not depend on $\boldmathsymbol{\Phi}$, but only on the location of the surfaces $\mathcal{C}_\uend$ and $\mathcal{C}_{\mathrm{uv}}$.

Following \Fig{fig:contour_poles}, the integration over the real axis $\gamma_0$ can be complemented by an integral over $\gamma_-$, which, thanks to the term $\ee^{-i t \N}$ in \Eq{eq:pdf:chi}, asymptotically vanishes (assuming that $g$, which appears in \Eq{eq:chi:pole:expansion}, does not increase exponentially or faster at large $\vert t\vert$). This leads to
\bea
 \label{eq:pdf_inegral_contour}
P_\boldmathsymbol{\Phi}(\N)=\sum_{n}\,a_n(\boldmathsymbol{\Phi})\,e^{-\Lambda_n\,\N}\,.
\eea
This form is always valid, but at large $\N$, only the first terms in the sum dominate, and it provides a tail expansion in terms of decaying exponentials. The dominant term is given by the lowest pole of the characteristic function $\Lambda_0$ and its residue $a_0(\boldmathsymbol{\Phi})$. In practice, the decay rates $\Lambda_n $ can be found by solving the characteristic function from \Eq{eq:diff:chi} and finding the zeros of its inverse. The residues $a_n(\boldmathsymbol{\Phi})$ can then be obtained from evaluating the derivative of the inverse characteristic function at $t=-i\Lambda_0$, \ie
\begin{align}
\label{eq:alphan:flat:generic}
a_n(\phi) &= -i \left[\frac{\partial}{\partial t}\chi_\N^{-1}\left(t=-i\Lambda_n,\phi\right)\right]^{-1}\, .
\end{align}
Let us stress again that, while the amplitude of the tail, controlled by $a_0$, depends on the initial field value, the decay rates $\Lambda_n$ are universal for a given potential. 
\subsection{An equivalent eigenvalue problem}
\label{sec:eigenvalues}
Let us now present an alternative method that leads to the same tail expansion, but that can be of complementary practical convenience. This relies on viewing \Eq{eq:adjoint:FP} as a heat equation, and employing well-known late-time limit techniques designed for heat or diffusion equations to solve it. Formally, \Eq{eq:adjoint:FP} can be solved as
\bea
\label{eq:adjoint:FokkerPlanck:formal:solution}
P_\boldmathsymbol{\Phi}\left(\mathcal{N}\right) = \exp\left[\mathcal{N}\mathcal{L}^\dagger_{\mathrm{FP}}\left(\boldmathsymbol{\Phi}\right)\right] P_\boldmathsymbol{\Phi}\left(\mathcal{N}=0\right)\, .
\eea

It is then important to notice that, although $\mathcal{L}^\dagger_{\mathrm{FP}}$ is not self-adjoint with respect to the scalar product defined below \Eq{eq:adjoint:FP}, one can introduce another scalar product, namely
\bea
\langle F_1, F_2 \rangle_w = \int \dd \boldmathsymbol{\Phi} F_1(\boldmathsymbol{\Phi}) F_2(\boldmathsymbol{\Phi}) w (\boldmathsymbol{\Phi})\, ,
\quad\quad\mathrm{where}\quad\quad w(\boldmathsymbol{\Phi}) = \frac{e^{1/v(\boldmathsymbol{\Phi})}}{v(\boldmathsymbol{\Phi})}\, ,
\eea
such that $\mathcal{L}^\dagger_{\mathrm{FP}}$ is self-adjoint with respect to that scalar product.\footnote{We again thank  Nahid Ahmadi, Zahra Ahmadi, Niloufar Feyz, Mahdiyar Noorbala and Borna Salehian for point that out to us.\label{footnote:ack2}} One can indeed show explicitly by integrating by parts that $(\mathcal{L}^\dagger_{\mathrm{FP}})^{\dagger_w}=\mathcal{L}^\dagger_{\mathrm{FP}}$ [while $(\mathcal{L}^\dagger_{\mathrm{FP}})^{\dagger}=\mathcal{L}_{\mathrm{FP}}\neq \mathcal{L}^\dagger_{\mathrm{FP}}$]. Therefore, the eigenvalues $-\Lambda_n$ of $\mathcal{L}^\dagger_{\mathrm{FP}}$ are all real (here a minus sign is introduced for notational convenience), and one can introduce an orthonormal set of eigenfunctions $\Psi_n$ of the operator $\mathcal{L}^\dagger_{\mathrm{FP}}$, 
\bea
\label{eq:eigen:value:problem}
\mathcal{L}^\dagger_{\mathrm{FP}}\cdot \Psi_n\left(\boldmathsymbol{\Phi}\right) = - \Lambda_n \Psi_n\left(\boldmathsymbol{\Phi}\right)\, ,
\eea
with boundary conditions $\Psi_n(\boldmathsymbol{\Phi})=0$ when $\boldmathsymbol{\Phi} \in \mathcal{C}_\uend$, and $[\bm{u}(\boldmathsymbol{\Phi})\cdot\bm{\nabla}]\Psi_n(\boldmathsymbol{\Phi})=0$ when $\boldmathsymbol{\Phi} \in \mathcal{C}_{\mathrm{uv}}$. Decomposing $P_\boldmathsymbol{\Phi} \left(\mathcal{N}=0\right)$ on the basis formed by these functions,
\bea
\label{eq:EigenProblem}
P_\boldmathsymbol{\Phi}\left(\mathcal{N}=0\right) = \sum_n \alpha_n \Psi_n\left(\boldmathsymbol{\Phi}\right)\, ,
\eea
\Eq{eq:adjoint:FokkerPlanck:formal:solution} gives rise to
\bea
\label{eq:tail_expansion:2}
P_\boldmathsymbol{\Phi}\left(\mathcal{N}\right) = \sum_n \alpha_n \Psi_n\left(\boldmathsymbol{\Phi}\right) \ee^{-\Lambda_n \mathcal{N}}\, .
\eea
This expression is nothing but the tail expansion~(\ref{eq:tail_expansion}), if one identifies $a_n(\boldmathsymbol{\Phi})=\alpha_n\Psi_n(\boldmathsymbol{\Phi})$.
It also shows that the characteristic function does not possess multiple pole, as announced above, otherwise the PDF would contain terms of the form $\mathcal{N}^{k-1}\ee^{-\Lambda_n \mathcal{N}}$, where $k$ is the order of the pole.

Let us note that the first boundary condition given below \Eq{eq:eigen:value:problem} comes from the requirement that $P_{\boldmathsymbol{\Phi}}(\N) = \delta(\N)$ when $\boldmathsymbol{\Phi} \in \mathcal{C}_\uend$, so all eigen-components should be identically zero for $\boldmathsymbol{\Phi} \in \mathcal{C}_\uend$ except when $\Lambda_n=\infty$. The second boundary condition simply comes from the reflective surface located at $\mathcal{C}_{\mathrm{uv}}$. 

One can also notice that \Eq{eq:eigen:value:problem} for the eigenfunctions $\Psi_n$ is the same as \Eq{eq:diff:chi} for the characteristic function, if one identifies $t$ with $-i\Lambda_n$. However, the boundary conditions are different, which makes the two problems technically different (and one can be more convenient to solve than the other), although perfectly equivalent. In particular, solving one problem automatically gives the solution for the other. Indeed, if \Eq{eq:diff:chi} has been solved and the functions $a_n(\boldmathsymbol{\Phi})$ derived, the coefficients $\alpha_n$ can be obtained as follows. Making use of the fact that the eigenfunctions $\Psi_n$ form an orthonormal set, \ie
\bea
\label{eq:Psin:normalisation}
\left\langle \Psi_n , \Psi_m \right\rangle_{w} =
\int_{\mathcal{C}} \Psi_n(\boldmathsymbol{\Phi}) \Psi_m(\boldmathsymbol{\Phi}) w(\boldmathsymbol{\Phi}) \dd \boldmathsymbol{\Phi} = \delta_{n,m},
\eea
where $\mathcal{C}$ is the field-space domain located between $\mathcal{C}_\uend$ and $\mathcal{C}_{\mathrm{uv}}$, \Eq{eq:tail_expansion} leads to
\bea
\left\langle \Psi_n ,P_\boldmathsymbol{\Phi}\left(\mathcal{N}\right)\right\rangle_{w} = \left[
\int_{\mathcal{C}} \Psi_n\left(\boldmathsymbol{\Phi}\right)
a_n(\boldmathsymbol{\Phi})
w(\boldmathsymbol{\Phi})
\dd\boldmathsymbol{\Phi}
\right]\,e^{-\Lambda_n\, \N}\, ,
\eea
while \Eq{eq:tail_expansion:2} gives rise to
\bea
\label{eq:flat:<Phi,P>:1}
\left\langle \Psi_n ,P_\boldmathsymbol{\Phi}\left(\mathcal{N}\right)\right\rangle_{w} = \alpha_n  \ee^{-\Lambda_n \mathcal{N}}\, .
\eea
By identifying the two expressions, one obtains
\bea
\label{eq:alphan:from:an}
\alpha_n = 
\int_{\mathcal{C}} \Psi_n\left(\boldmathsymbol{\Phi}\right)
a_n(\boldmathsymbol{\Phi})
w(\boldmathsymbol{\Phi})
\dd\boldmathsymbol{\Phi}\, .
\eea
Conversely, if the decomposition~(\ref{eq:EigenProblem}) has been performed and the coefficients $\alpha_n$ are known, the functions $a_n(\boldmathsymbol{\Phi})$ can be obtained from the relation $a_n(\boldmathsymbol{\Phi})=\alpha_n\Psi_n(\boldmathsymbol{\Phi})$ given below \Eq{eq:tail_expansion:2}.

\section{Applications}
\label{sec:applications}
We now apply the techniques developed in the previous section to concrete examples. We investigate four single-field, slow-roll inflationary potentials: an exactly flat potential in \Sec{sec:flat_potential}, a potential with a constant slope in \Sec{sec:linear_potential}, a potential with a cubic flat inflection point in \Sec{sec:inflection_potential}, and a potential with a linearly-tilted cubic inflection point in \Sec{sec:inflection_linear_potential}. The adjoint Fokker-Planck operator is given by \Eq{eq:FP:SR}, and the boundary conditions simply consist in an absorbing wall located at $\phi_\uend$ and a reflective wall located at $\phiuv$.
\subsection{Flat potentials}
\label{sec:flat_potential}
\begin{figure}[t]
\centering 
\includegraphics[width=.49\textwidth]{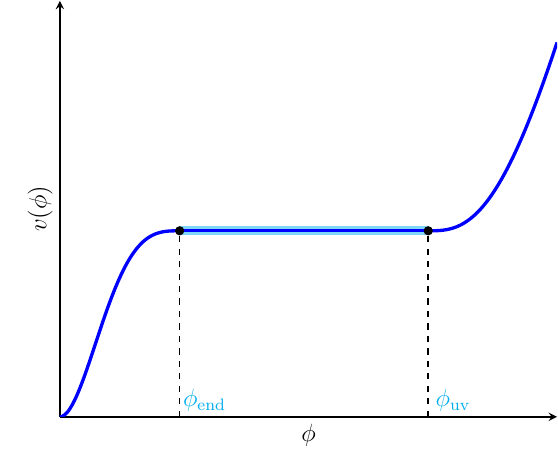}
\includegraphics[width=.49\textwidth]{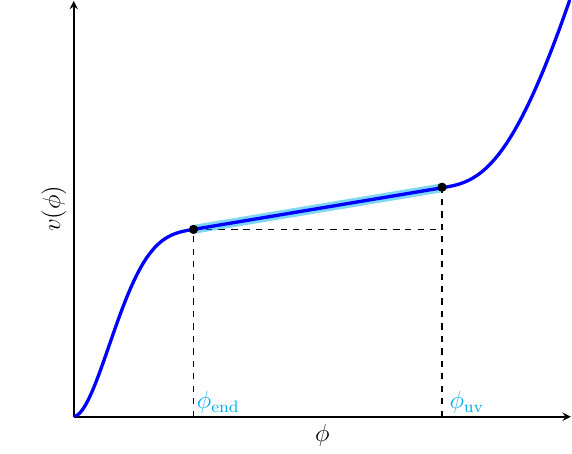}
 \caption{Schematic representation of the flat (left) and linear (right) potentials studied in sections \ref{sec:flat_potential} and \ref{sec:linear_potential} respectively. We only consider the region between $\phiend$ and $\phiuv$.}
 \label{fig:flat_linear_potential}
\end{figure}
Let us begin by considering a potential that is exactly flat, $v=v_0$, between $\phiend = 0$ and $\phiuv=\dphiwell$, as in the left panel of \Fig{fig:flat_linear_potential}. In principle, if the potential is exactly flat, slow roll is violated since there is no potential gradient. However, in \Ref{Pattison:2017mbe}, it was shown that for a potential of the form $v = v_0 [1+(\phi/\phi_0)^p]$, where $\phi_0\gg \Mp$ such that slow roll is never violated, in the region of the potential located between $\phi=0$ and $\phi=\dphiwell = \phi_0 v_0^{1/p}$, the potential gradient term  in \Eq{eq:FP:SR} (\ie the first term on the right-hand side) can be neglected, and the classical part of the potential above $\dphiwell$ acts as a reflective wall (see also the discussion around \Eq{eq:classicality_criterion} below). The full results were thus shown to be very accurately reproduced if one places a reflective boundary condition at $\phi=\dphiwell $ and considers a pure diffusion process between $\phi=0$ and $\phi=\dphiwell$. 

This is the situation we consider here, where ``flat'' potential has to be taken in the sense of that specific limit, in which slow roll is not violated. The problem was entirely solved in \Ref{Pattison:2017mbe}, and we want to check the consistency of our approach with their results.
\subsubsection{Poles of the characteristic function.}
\begin{figure}[t!]
\centering 
\includegraphics[width=.59\textwidth]{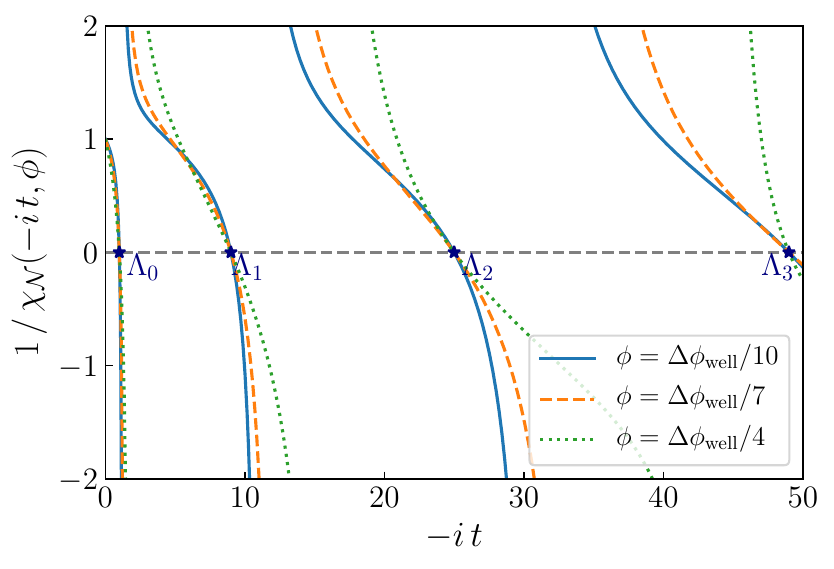}
\caption{Zeros of the inverse characteristic function for a flat potential. We have chosen $\mu^2=\pi^2/4$ so that zeros are located at $\Lambda_n=(2n+1)^2$. We have evaluated $\chi^{-1}_\N(t,\phi)$ at different field values $\phi$. Although the characteristic functions for each $\phi$ are different, the location of the poles $\Lambda_n$ (which determine the decay rates of the PDF) is universal for a given potential.}
 \label{fig:ichi_flat}
\end{figure}

In this simple example, the equation for the characteristic function \eqref{eq:diff:chi} reads
\bea
\label{eq:eom:chi:flat}
\chi''_\N(t,\phi) + \frac{i\,t}{v_0\Mp^2} \chi_\N(t,\phi) = 0\,,
\eea
where a prime denotes derivation with respect to the field value $\phi$, with boundary conditions $\chi_\N(t,0)=1$ and $\chi_\N^\prime(t,\phiuv)=0$. It can be solved as
\bea 
\label{eq:chi_flat}
\chi_\N(t,\phi)=\frac{\cos\lb(it)^{1/2}\mu(x-1)\rb}{\cos\lb(it)^{1/2}\mu\rb}\,,
\eea
where we have defined $x=\phi/\dphiwell$ and introduced the quantity
\bea \label{eq:mu}
\mu^2 = \frac{\dphiwell^2}{v_0\Mp^2}\,
\eea
as in in \Ref{Pattison:2017mbe}. The poles of \Eq{eq:chi_flat} correspond to when the argument of the $\cos$ function in the denominator equals $(n+1/2)\pi$, where $n$ is an integer number, and calling $\Lambda_n$ the value of $it$ at these poles, one has
\bea \label{eq:flat_poles}
\Lambda_n=\frac{\pi^2}{\mu^2}\lp n+\frac{1}{2}\rp^2\,.
\eea
One can check that, in agreement with the discussion of \Sec{sec:pole:chi}, the $\Lambda_n$'s are all real, positive and independent of $\phi$.
The exponential decay rate of the tail of the PDF therefore depends both on the width of the quantum well, $\dphiwell$, and its scale $v_0$, through the combination $\mu$. We have plotted the inverse characteristic function for a flat potential in \Fig{fig:ichi_flat}, for a few field values, where the zeros of $\chi^{-1}_\N(t,\phi)$ correspond to $-i\Lambda_n$. This illustrates again that, although the details of the characteristic functions depend on $\phi$, the location of their poles $\Lambda_n$ is universal for a given potential.

Finally, making use of \Eq{eq:alphan:flat:generic}, the coefficients $a_n$ are given by
\begin{align}
a_n(\phi) = (-1)^n \frac{\pi}{\mu^2}\left(2n+1\right) \cos\left[\frac{\pi}{2}\left(2n+1\right)\left(x-1\right)\right] .
\label{eq:an:flat}
\end{align}

\subsubsection{Eigenvalue problem.}
In the case of a flat potential, the eigenvalue problem \eqref{eq:eigen:value:problem} reads
\bea
\label{eq:Psin:eom:flat}
\Psi_n''\left(\phi\right) + \frac{\Lambda_n}{v_0\Mp^2} \Psi_n\left(\phi\right) = 0,
\eea
with boundary conditions $\Psi_n(0) = \Psi_n'(\dphiwell)=0$.  
The generic solution of \Eq{eq:Psin:eom:flat} is
\bea
\Phi_n\left(\phi\right) = A_n \exp\left(i\sqrt{\frac{\Lambda_n}{v_0\Mp^2}}\phi\right) + B_n \exp\left(-i\sqrt{\frac{\Lambda_n}{v_0\Mp^2}}\phi\right)\, ,
\eea
where the first boundary condition imposes that $B_n=-A_n$, hence $\Psi_n(\phi) \propto \sin[\sqrt{{\Lambda_n}/({v_0\Mp^2})}\phi]$. The second boundary condition then implies that $\cos[\sqrt{{\Lambda_n}/({v_0\Mp^2})}\dphiwell]=0$, which precisely gives rise to \Eq{eq:flat_poles}. Normalising the functions $\Psi_n$ as in \Eq{eq:Psin:normalisation}, one then has
\bea
\label{eq:phin_flat}
\Psi_n\left(\phi\right)= \sqrt{\frac{2 v_0}{\dphiwell \ee^{1/v_0}}} \sin\left[\pi\left(n+\frac{1}{2}\right)\frac{\phi}{\dphiwell}\right]
 .
\eea
The coefficients $\alpha_n$ can be computed from \Eq{eq:alphan:from:an}, and \Eqs{eq:an:flat} and~(\ref{eq:phin_flat}) give rise to
\bea
\label{eq:alphan:flat}
\alpha_n = \frac{2\pi}{\mu^2}\left(n+\frac{1}{2}\right) \sqrt{\frac{\dphiwell \ee^{1/v_0}}{2v_0}}\, .
\eea
Altogether, the PDF for the constant potential can be written as 
\bea 
\label{eq:flat_PDF}
P_\phi(\N)= \frac{2\pi}{\mu^2} \sum_{n=0}^\infty\left(n+\frac{1}{2}\right) \, \sin \left[\pi\left(n+\frac{1}{2}\right)\frac{\phi}{\phiuv}\right] \, \ee^{-\left(n+\frac{1}{2}\right)^2 \frac{\pi^2}{\mu^2}\,\N}\, ,
\eea
which exactly matches Eq.~(4.10) of \Ref{Pattison:2017mbe}, and which, as noted there, can be expressed in terms of the derivative of an elliptic theta function
\bea
\label{eq:flat_PDF:elliptic}
P_\phi(\N)= -\frac{\pi}{2\mu^2}\vartheta_2'\left(\frac{\pi\phi}{2\phiuv},\ee^{-\frac{\pi^2}{\mu^2}\N}\right)\, .
\eea

\begin{figure}[t!]
\centering 
\includegraphics[width=.59\textwidth]{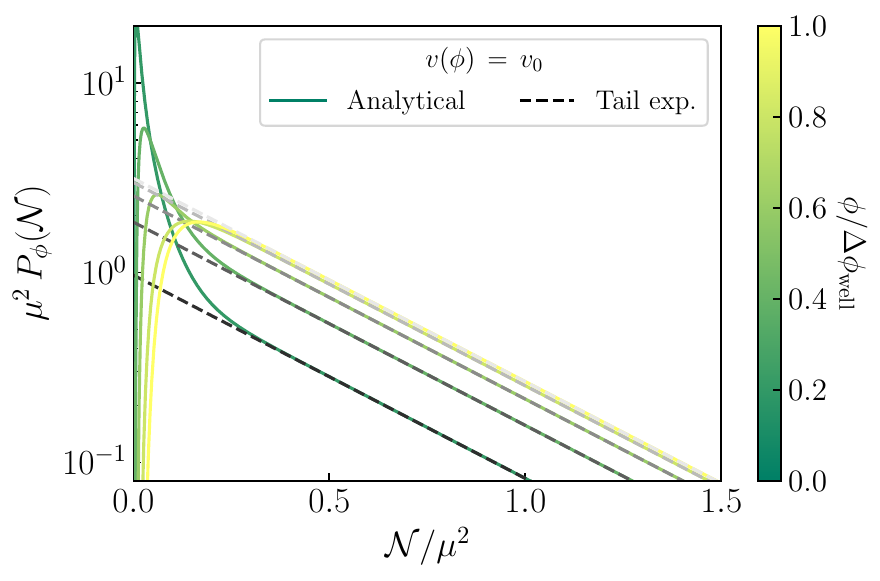}
 \caption{Probability distribution function of the number of \efolds~$\N$ in a flat potential, starting from different initial field values $\phi$. We compare the full PDF, \Eq{eq:flat_PDF} (solid lines), with the leading term in the tail expansion~(\ref{eq:pdf_inegral_contour}) (dashed lines). We rescale the axes by $\mu^2=\dphiwell ^2/(v_0 \Mp^2)$, such that, using the self-similarity of \Eq{eq:flat_PDF}, the result does not depend on~$\mu$.} 
 \label{fig:PDF_flat}
\end{figure}

In \Fig{fig:PDF_flat}, we plot both the full PDF~(\ref{eq:flat_PDF}) and the leading term in the tail expansion~(\ref{eq:pdf_inegral_contour}), $a_0(\phi)e^{-\Lambda_0\N}$. As it can be observed, at large-$\N$ values, the tail expansion provides an excellent approximation to the full result. Note also that \Eq{eq:flat_PDF} is such that the PDF of the quantity $\N/\mu^2$ is independent of $\mu$, which is why this quantity is displayed in \Fig{fig:PDF_flat}. This shows that increasing $v_0$, or decreasing $\dphiwell$, decreases the typical values of $\N$.

\subsection{Potentials with constant slope}
\label{sec:linear_potential}

Let us now consider a potential of the type
\bea
\label{eq:pot:constant:slope}
v\left(\phi\right) = v_0\left(1+\alpha\frac{\phi}{\Mp}\right),
\eea
with a constant slope $\alpha$, which we will assume is positive without loss of generality. The model~\eqref{eq:pot:constant:slope} is bounded between $\phi=0$ where the potential is supposed to become steeper and/or inflation ends; and $\phi=\phiuv$ where the potential is supposed to become steeper and the dynamics of $\phi$ dominated by classical drift, which acts as a reflective wall as discussed at the beginning of \Sec{sec:flat_potential}. Since $\epsilon_1\simeq \alpha^2/2$, one should have $\alpha\ll 1$ in order for slow roll to be valid. Moreover, we will consider only scenarios where $\phi_{\mathrm{uv}}\ll \Mp/\alpha$, such that the potential is almost constant, $v\simeq v_0$, between $\phi=0$ and $\phi=\phiuv$ (see the right panel of \Fig{fig:flat_linear_potential}). 

\subsubsection{Poles of the characteristic function.}
The equation for the characteristic function \eqref{eq:diff:chi} is given by
\bea \label{eq:chi_linear_general}
\chi_\N''(t,\phi) - \frac{v_0\alpha}{\Mp v(\phi)^2}\chi_\N'(t,\phi)+ \frac{i\,t}{\Mp^2v(\phi)} \chi_\N(t,\phi) = 0\,,
\eea
which, compared to \Eq{eq:eom:chi:flat} for a flat potential, contains an additional friction term. The other difference is that now, the coefficients of the differential equation depend on $\phi$, so there is no generic analytic solution. 
There is, however, an analytic solution in the ``almost-constant'' regime, $\phi_{\mathrm{uv}}\ll \Mp/\alpha$, where $v(\phi)$ 
and all its derivatives are replaced by their leading-order expression in $\alpha$.
 This solution reads
\bea
\label{eq:chiN:cosh:ext}
\chi_\N\left(t, \phi \right) = \ee^{\frac{\alpha \mu x}{2  \sqrt{v_0}}}\frac{2\gamma\sqrt{it v_0} \cos \left[ \mu \gamma \sqrt{it}(x-1)\right] - \alpha  \sin \left[ \mu \gamma \sqrt{it}(x-1)\right]}{2\gamma \sqrt{ i t v_0} \cos \left({\mu} \gamma\sqrt{it }\right)+\alpha  \sin \left({\mu} \gamma\sqrt{it }\right)}\, ,
\eea
with 
\bea
\gamma = \sqrt{1-\frac{\alpha^2}{4itv_0}}\, ,
\eea
$x=\phi/\phiuv$, and $\mu$ is given by \Eq{eq:mu} where $\dphiwell$ is replaced by $\phiuv$. When $\alpha=0$, this boils down to the flat potential solution~\eqref{eq:chi_flat}.

\begin{figure}[t!]
\centering 
\includegraphics[width=.59\textwidth]{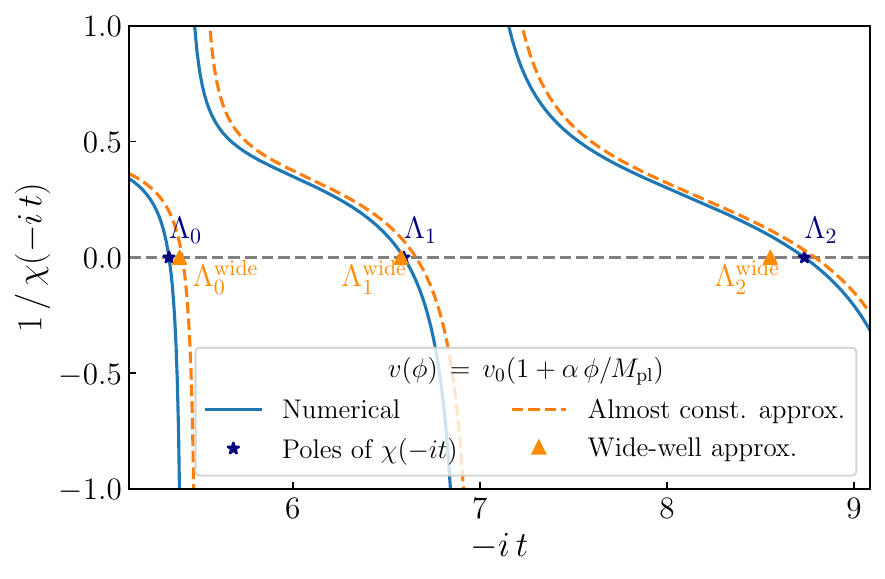}
\caption{Inverse characteristic function for a linear potential. We compare the numerical solution of \Eq{eq:chi_linear_general} (solid blue line) and its poles $\Lambda_n$ (labeled with blue stars) with the ``almost-constant'' approximation~\eqref{eq:chiN:cosh:ext} (orange dashed line) and its approximate poles $\Lambda^{\mathrm{wide}}_n$ in the ``wide-well approximation''~\eqref{eq:lambda_almost_const} (labeled with orange triangles). 
These two approximations are valid when $\phiuv/\Mp\ll1/\alpha$, and $\phiuv/\Mp\gg v_0/\alpha$, respectively. 
The characteristic function is evaluated at $\phi=\phiuv/10$, with $\alpha=0.1$, and $v_0$ and $\phiuv$ have been set such that $\phiuv/\Mp=0.01/\alpha$ and $\phiuv/\Mp = 20 v_0/\alpha$. Increasing $1/\alpha$ and decreasing $v_0/\alpha$ improves the agreement between the three results, but quickly makes the separation between the $\Lambda_n$ impossible to resolve by eye, which is why somewhat intermediate values have been used here, for illustrative purpose.
}
 \label{fig:ichi_linear}
\end{figure}

The poles of the characteristic function at $t=-i\Lambda_n$ are determined by the equation
\bea \label{eq:trascendental_linear}
\tan\left(\sqrt{\Lambda_n-\frac{\alpha^2}{4v_0}}\mu\right)=- 2   \frac{v_0}{\alpha}\frac{\Mp}{\phiuv}     \sqrt{\Lambda_n-\frac{\alpha^2}{4v_0}}\mu\, .
\eea
This equation is of the form $\tan(z)=-2 a z$, with $z=\sqrt{\Lambda_n-\alpha^2/(4v_0)}\mu$ and $a=v_0\Mp/(\alpha\phiuv)$. It has one obvious solution, namely $z=0$, which would lead to $\Lambda=\alpha^2/(4 v_0)$. However, the numerator of \Eq{eq:chiN:cosh:ext} also vanishes at $t=-i \alpha^2/(4 v_0)$, and by carefully expanding the characteristic function around that value, one can see that it is in fact regular, and does not possess a pole. The case $z=0$ can therefore be safely discarded, and for $z>0$, one has to solve a transcendental equation, which cannot be done analytically. However, approximate solutions can be found in the two limits $a\ll 1$ and $a\gg 1$, which we dub the ``wide-well'' and the ``narrow-well'' regimes respectively, since they imply a lower bound and an upper bound on $\phiuv$ respectively.
\paragraph{Wide-well limit $\phiuv/\Mp \gg v_0/\alpha$}
In this case $a\ll 1$, hence $\vert \tan(z)/z \vert \ll 1$, which implies that $z$ is close to $(n+1)\pi$, with $n$ an integer number. One can write $z=(n+1)\pi+\delta z$, and expand $\tan(z)\simeq \delta z + \delta z^3/3+\cdots$. Plugging this formula into the transcendental equation, and expanding in $a$, one obtains $\delta z\simeq -2 a (n+1)\pi[1-2a (n+1) \pi + \cdots]$, which gives rise to
\bea
\label{eq:lambda_almost_const}
\Lambda_n^\mathrm{wide} = \frac{\alpha^2}{4 v_0}+ \frac{\left(n+1\right)^2 \pi^2}{\mu^2}\left(1-4 \frac{v_0 \Mp}{\alpha\phiuv} + \cdots \right)
\eea
where ``$\cdots$'' denotes higher powers of $v_0\Mp/(\alpha\phiuv)$, so this approximation is indeed valid in the regime
\bea
\label{eq:constant:slope:expansion:transcendental:condition}
\frac{\phiuv}{\Mp} \gg \frac{v_0}{\alpha} \, .
\eea
If one takes $\phiuv$ to its maximal allowed value, $\phiuv\sim \Mp/\alpha$, this condition is satisfied as soon as $v_0\ll 1$, which is always the case. 

\begin{figure}[t!]
\centering 
\includegraphics[width=.59\textwidth]{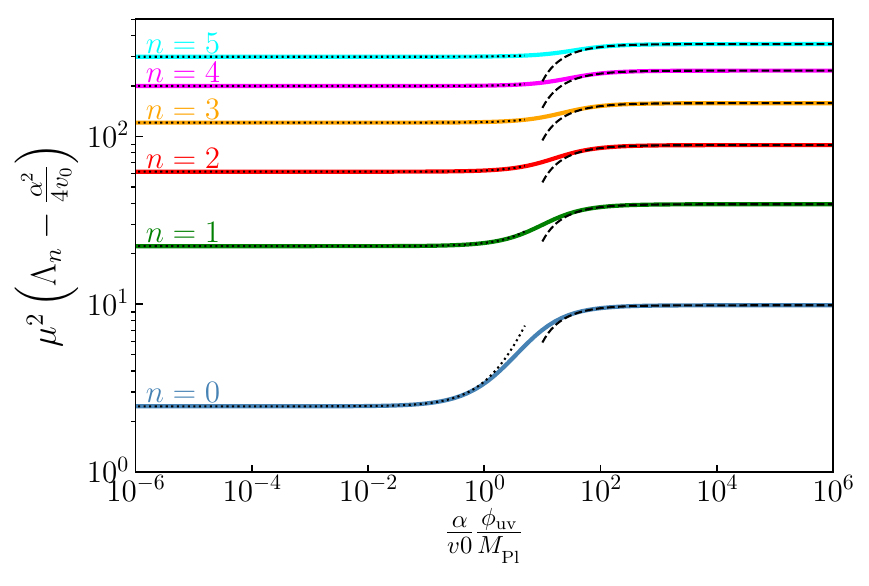}
\caption{Decay rates $\Lambda_n$ for the constant-slope potential~\eqref{eq:pot:constant:slope}, as a function of $\alpha\phiuv/(v_0\Mp)$. The coloured lines correspond to numerical solutions of the transcendental equation~\eqref{eq:trascendental_linear}, the black dashed lines display the wide-well approximation~\eqref{eq:lambda_almost_const}, and the black dotted lines stand for the narrow-well approximation~\eqref{eq:lambda_almost_const:narrow}.}
 \label{fig:constantslope:Lambdan}
\end{figure}

In \Fig{fig:ichi_linear} we show the inverse characteristic function obtained by solving numerically \Eq{eq:chi_linear_general} (solid blue line) and the analytical solution~\eqref{eq:chiN:cosh:ext} in the almost-constant approximation (orange dashed line). We also include the approximate values of $\Lambda_n^\mathrm{wide}$ in the wide-well limit given in \Eq{eq:lambda_almost_const}. In order to remain in the regime of validity of this approximation but to make visible the differences between these different estimates, we set the parameters such that $\phiuv/\Mp=0.01/\alpha$ and $\phiuv/\Mp = 20 v_0/\alpha$. By increasing $1/\alpha$ and decreasing $v_0/\alpha$, the agreement between the three results largely improves, but quickly makes the separation between the $\Lambda_n$ impossible to resolve by eye, which is why intermediate values have been used here, for illustrative purpose. 
To test further the consistency of these results, the formula~\eqref{eq:lambda_almost_const} is compared with a numerical solution of the transcendental equation~\eqref{eq:trascendental_linear} in \Fig{fig:constantslope:Lambdan}, where one can check that, as long as \Eq{eq:constant:slope:expansion:transcendental:condition} is valid, it provides indeed a good approximation.

By comparing \Eqs{eq:flat_poles} and~(\ref{eq:lambda_almost_const}), one can check that, as in the flat case, $\Lambda_n$ receives a $n$-dependent contribution proportional to $\pi^2/\mu^2$, but it is also shifted by a fixed quantity, namely $\alpha^2/(4 v_0)$, which dominates over the $n$-dependent contribution, because of the condition~(\ref{eq:constant:slope:expansion:transcendental:condition}). One concludes that, in the wide-well regime, adding a small slope to the potential is enough to highly suppress the tails.

Finally, the $a_n$ functions can be approximated as follows. One needs to expand \Eq{eq:chiN:cosh:ext} around the poles $t=-i\Lambda_n$ in order to extract the residues. Since the decay rates $\Lambda_n$ are not known exactly, this expansion cannot be done directly. However, writing $\Lambda_n = \Lambda_n^{(0)} + \delta \Lambda_n $, where $\Lambda_n^{(0)}$ corresponds to the approximation~(\ref{eq:lambda_almost_const}), one can parametrise $t$ in the neighbourhood of the poles as $t=- i  \Lambda_n^{(0)}  - i  \delta \Lambda_n + \delta t$, and expand the characteristic function in $\delta t$. Obviously, the function cannot be probed on scales smaller than $\delta \Lambda_n$, so one assumes in fact $\delta \Lambda_n \ll \delta t \ll  \Lambda_n^{(0)}$, and performs a double expansion in $ \delta \Lambda_n$ and $\delta t$ under these conditions. By identification with \Eq{eq:chi:pole:expansion}, the residues can then be extracted, and one obtains
\begin{align}
a_n^{\mathrm{wide}}(\phi) = -(-1)^n \frac{\pi}{\mu^2}2\left(n+1\right) e^{\frac{\alpha\phiuv}{2v_0\Mp} x} \sin\left[{\pi}\left(n+1\right)\left(x-1\right)\right] .
\label{eq:a:constant:slope}
\end{align}
One notices that the structure is similar to, though different from, the one for a flat potential~\eqref{eq:an:flat}. In particular, the exponential term gives a strong enhancement, because of \Eq{eq:constant:slope:expansion:transcendental:condition}.
\paragraph{Narrow-well limit $\phiuv/\Mp \ll v_0/\alpha$}
In the opposite limit where $a\gg 1$, $\vert \tan(z)/z \vert \gg 1$, which implies that $z$ must be close to $\pi/2+n\pi$, where $n$ is an integer number. One can write $z=\pi/2+n\pi+\delta z$, and expand $\tan(z)\simeq -1/\delta z+\delta z/3+\cdots$. Plugging this formula into the transcendental equation, and expanding in powers of $1/a$, one obtains $\delta z \simeq 1/[2\pi a(n+1/2)]+\cdots$, which gives rise to
\bea
\label{eq:lambda_almost_const:narrow}
\Lambda_{n}^\mathrm{narrow} = \frac{\pi^2}{\mu^2}\left[\left(n+\frac{1}{2}\right)^2+\frac{\alpha\phiuv}{\pi^2 v_0\Mp}+\cdots\right],
\eea
where ``$\cdots$'' denotes higher powers in $\alpha\phiuv/(v_0\Mp)$, so this approximation indeed holds in the regime 
\bea
\label{eq:small:alpha:limit:def}
\frac{\phiuv}{\Mp} \ll \frac{v_0}{\alpha} \, .
\eea
The formula~\eqref{eq:lambda_almost_const:narrow} is compared with a numerical solution of \Eq{eq:trascendental_linear} in \Fig{fig:constantslope:Lambdan}, where one can check that, as long as \Eq{eq:small:alpha:limit:def} is valid, it indeed provides a good approximation.

By comparing \Eqs{eq:flat_poles} and~(\ref{eq:lambda_almost_const:narrow}), one can see that the difference with the flat-potential case is negligible: adding a slope in the narrow-well regime only shifts the spectrum by a correction, $\alpha\phiuv/(v_0\Mp)$, which is, by definition, tiny in that regime. The same procedure as the one outlined above \Eq{eq:alphan:constant:slope} can also be performed in order to extract the $a_n$ functions. At leading order in $\alpha\phiuv/(v_0\Mp)$, one exactly recovers \Eq{eq:alphan:flat}, which finishes to prove that the narrow-well regime in fact corresponds to the flat-potential limit. Since \Eq{eq:small:alpha:limit:def} can also be interpreted as an upper bound on $\alpha$, this result makes sense.
\paragraph{Comparison with the classical result}
The classical value of the power spectrum, $\calP_{\zeta,\mathrm{cl}}=2v^3/(\Mp^2v'^2)$, in this model, is given by $\calP_{\zeta,\mathrm{cl}}\simeq 2v_0/\alpha^2$. Since $\Lambda_0= \alpha^2/(4 v_0)$ in the wide-well regime, the dominant behaviour on the tail can thus be written as 
\bea
\label{eq:constant:slope:wide:well:tail:comp:class}
P_\phi^{\mathrm{wide}}(\mathcal{N}) \underset{\mathcal{N}\gg 1}{\propto} \exp\left(-\frac{1}{2}\frac{\mathcal{N}}{\calP_{\zeta,\mathrm{cl}}}\right)\, .
\eea
By comparison, in the classical picture, \Eq{eq:tail_expansion:classical} gives rise to
\bea
\label{eq:constant:slope:class:tail}
\left. P_\phi(\mathcal{N}) \right\vert_{\mathrm{cl}} 
\underset{\mathcal{N}\gg 1}{\propto}
 \exp\left(-\frac{1}{2}\frac{\mathcal{N}^2}{\calP_{\zeta,\mathrm{cl}} N_\ucl}\right)\, ,
\eea
where $N_\ucl$ is the classical number of \efolds~that arises from the integration over $k$ in \Eq{eq:tail_expansion:classical}, which can trivially be performed since $\calP_{\zeta,\mathrm{cl}}$ is independent of $\phi$ in the almost-constant approximation. Two remarks are in order. First, as mentioned above, as soon as $\N\gg N_\ucl$, the amount of power on the tail is greatly enhanced in the full stochastic theory compared to the classical, Gaussian approximation. Second, the similarity between \Eqs{eq:constant:slope:wide:well:tail:comp:class} and~(\ref{eq:constant:slope:class:tail}), which coincide for $\N=N_\ucl$, is an illustration of the resemblance between the classical and the full stochastic theory for a linear potential. Indeed, as shown in \Ref{Pattison:2017mbe}, the classical limit can be obtained from the full stochastic PDF by a saddle-point expansion, where higher-order corrections involve either $v$, which is always small, or derivatives of the potential of order 2 or higher, which vanish in the present case. For instance, the classical number of \efolds~is given by $N_\ucl = \int_{\phi_\uend}^\phi v/v' \dd \tilde{\phi}/\Mp^2$, which, in the almost-constant approximation, reduces to 
\bea
\label{eq:Ncl:constant:slope}
N_\ucl = \frac{\phi-\phi_\uend}{\alpha \Mp} .
\eea
In the full stochastic theory, by Taylor expanding the exponential function in the first equality of \Eq{eq:characteristicFunction:def}, one can see that the mean number of \efolds~can be obtained from differentiating the characteristic function with respect to $t$, and evaluating the result at $t=0$. Making use of \Eq{eq:chiN:cosh:ext}, one obtains
\begin{align}
\label{eq:mean:N:chi}
\left\langle \N \right\rangle &= 
-i \left. \frac{\partial\chi_\N (t,\phi)}{\partial t}\right\vert_{t=0}
\\ & =
N_\ucl + \frac{v_0}{\alpha^2} \left[ \ee^{-\frac{\alpha}{v_0\Mp}\left(\phiuv-\phi_\uend\right)}- \ee^{-\frac{\alpha}{v_0\Mp}\left(\phiuv-\phi\right)}\right] .
\label{eq:mean:N:constant:slope}
\end{align}
In the wide-well regime, \ie when the condition~(\ref{eq:constant:slope:expansion:transcendental:condition}) is satisfied, the second term in the above expression is exponentially suppressed (unless one starts at a value of $\phi$ very close to $\phiuv$), and 
\bea
\langle \N \rangle_{\mathrm{wide}} \simeq N_\ucl .
\eea
In the narrow-well regime however, the effect of the boundary located at $\phiuv$ is not negligible anymore, and  expanding \Eq{eq:mean:N:constant:slope} in the limit~(\ref{eq:small:alpha:limit:def}), one finds
\bea
\left\langle \mathcal{N} \right\rangle_\mathrm{narrow} \simeq \frac{\left(\phi-\phi_\uend\right)\left(2\phiuv-\phi-\phi_\uend\right)}{2\Mp^2 v_0},
\eea
which is very different from, and in fact much smaller than, its classical counterpart~\eqref{eq:Ncl:constant:slope}.

\subsubsection{Eigenvalue problem} 
The eigenvalue problem is analogous to solving the equation for $\chi_\N(t,\phi)$, \Eq{eq:chi_linear_general}; there is no general solution. In the almost-constant approximation,
\bea \label{eq:eigenequation_linear}
\Psi_n''\left(\phi\right) - \frac{\alpha}{\Mp v_0}\Psi_n'\left(\phi\right)+ \frac{\Lambda_n}{v_0\Mp^2} \Psi_n\left(\phi\right) = 0\,,
\eea
the solution reads
\bea
\displaystyle
\Psi_n\left(\phi\right) = \ee^{\frac{\alpha\phi}{2\Mp v_0}}\left(A_n \exp^{i\sqrt{v_0 \Lambda_n - \frac{\alpha^2}{4}}\frac{\phi}{v_0\Mp}}+B_n \ee^{-i\sqrt{v_0\Lambda_n - \frac{\alpha^2}{4}}\frac{\phi}{v_0\Mp}}\right) .
\eea
The first boundary condition imposes $B_n=-A_n$, and the second one implies the transcendental equation \eqref{eq:trascendental_linear}, which has solution~\eqref{eq:lambda_almost_const} in the wide-well limit~\eqref{eq:constant:slope:expansion:transcendental:condition}, and solution~\eqref{eq:lambda_almost_const:narrow} in the narrow-well limit~\eqref{eq:small:alpha:limit:def}. The eigenfunctions are thus given by
\bea
\label{eq:constant:slope:vaccum:dom:Phin}
\Psi_n\left(\phi\right)=\sqrt{\frac{2 v_0}{\phiuv \ee^{1/v_0}}}\exp\left(\frac{\alpha\phi}{2\Mp v_0}\right)\sin\left(\sqrt{v_0 \Lambda_n - \frac{\alpha^2}{4}}\frac{\phi}{v_0\Mp}\right)\, ,
\eea
which boils down to \Eq{eq:phin_flat} if $\alpha=0$, and where the eigenfunctions have been normalised in the (extended) sense that $\langle \Phi_n^{(-\alpha)} , \Phi_m^{(\alpha)}\rangle_{w} = \delta_{n,m}$ (and using the ``almost-constant'' approximation to evaluate the kernel $w$, for consistency). 

The coefficients in the expansion~(\ref{eq:tail_expansion}), $\alpha_n$, can be determined by following the procedure outlined at the end of \Sec{sec:eigenvalues}, where in \Eq{eq:alphan:from:an}, our extended scalar product has to be used, \ie $ \alpha_n = \langle \Psi_n^{(-\alpha)}(\phi) , a_n^{(\alpha)}(\phi)\rangle $. In the wide-well regime, this leads to
\bea
\label{eq:alphan:constant:slope}
\alpha_n^\mathrm{wide} = \frac{2\pi}{\mu^2}\left(n+1\right) \sqrt{\frac{\phiuv \ee^{1/v_0}}{2 v_0}}\, .
\eea
Notice that it is similar to \Eq{eq:alphan:flat} for a flat potential, the only difference being that $n+1/2$ is replaced by $n+1$. Combining the above results, the PDF can be approximated as
\bea
P_\phi^{\mathrm{wide}}(\N) = 2\frac{\pi}{\mu^2} e^{\frac{\alpha \phiuv}{2 v0 \Mp}\frac{\phi}{\phiuv}} \sum_{n=0}^\infty n \sin\left(\pi n \frac{\phi}{\phiuv}\right)\ee^{-\left(\frac{\alpha^2}{4 v_0}+n^2 \frac{\pi^2}{\mu^2}\right)\N}\, .
\eea
One notices that the structure of the result is very similar to the one for a flat potential, \Eq{eq:flat_PDF}, if one replaces $n+1/2$ by $n$ in the sum, with the crucial difference that, now, the tails are suppressed by an additional $\ee^{-\alpha^2\N/(4 v_0)}$ factor. As for \Eq{eq:flat_PDF}, the result can be expressed in terms of the derivative of an elliptic theta function
\bea
\label{eq:constant:slope:PDF:elliptic}
P_\phi^\mathrm{wide}(\N)= -\frac{\pi}{2\mu^2}
\ee^{\frac{\alpha \phiuv}{2 v0 \Mp}\frac{\phi}{\phiuv}}
\ee^{-\frac{\alpha^2}{4 v_0}\N}
\vartheta_3'\left(\frac{\pi\phi}{2\phiuv},\ee^{-\frac{\pi^2}{\mu^2}\N}\right)\, .
\eea
In the narrow-well limit, as explained above, the flat-potential formulas are recovered, hence one obtains \Eq{eq:alphan:flat}, and the flat-potential PDF~\eqref{eq:flat_PDF} is obtained, up to small corrections suppressed by $\phiuv\alpha/(\Mp v_0)$.

Let us finally recall that the above results have been derived in the almost-constant approximation, which holds when $\phiuv\ll \Mp/\alpha$. To go beyond, one can employ the WKB approach outlined in \App{app:wkb}, in which a slightly different transcendental equation has to be solved.

In summary, for a potential with a constant slope $\alpha$ over a certain field range $\phiuv$, either the range is narrow in the sense that $\phiuv/\Mp\ll v_0/\alpha$, and the potential can be approximated as constant, such that the results of \Sec{sec:flat_potential} can be used; or the range is wide in the sense that $\phiuv/\Mp\gg v_0/\alpha$, and the PDF receives a strong suppression $\ee^{-\alpha^2\N/(4 v_0)}$ on its tail compared to the flat potential case.

\begin{figure}[t!]
\centering 
\includegraphics[width=.49\textwidth]{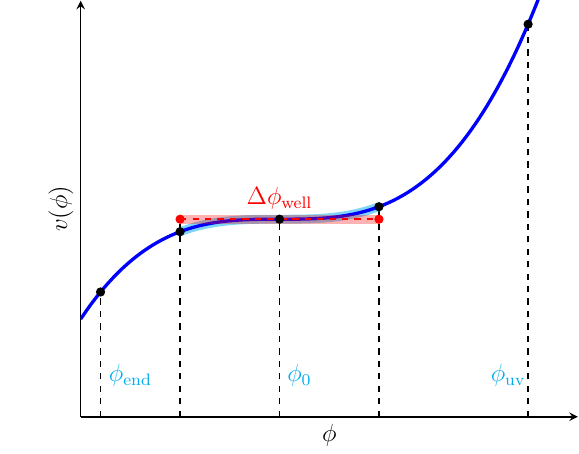}
\includegraphics[width=.49\textwidth]{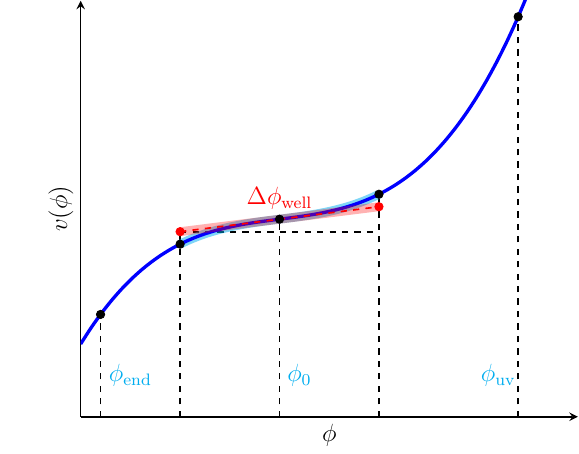}
 \caption{Schematic representation of the inflection (left) and tilted inflection (right) point potentials studied in \Secs{sec:inflection_potential} and \ref{sec:inflection_linear_potential} respectively. We solve the stochastic evolution between $\phiend$ and $\phiuv$. The region where quantum diffusion dominates, of width $\dphiwell$, can be approximated by a flat potential or a constant-slope potential respectively, and is determined by the non-classicality criterion (\ref{eq:classicality_criterion}).}
 \label{fig:inflection_linear_potential}
\end{figure}

\subsection{Inflection point potentials}
\label{sec:inflection_potential}

The toy models analysed in the two previous sections, the flat potential in \Sec{sec:flat_potential} and the constant-slope potential in \Sec{sec:linear_potential}, can serve as building blocks to study more realistic scenarios, that we now investigate in the two following sections. The first one is a potential with a flat inflection point located at $\phi_0$, as schematically displayed in the left panel of \Fig{fig:inflection_linear_potential}. In practice, we consider for simplicity a cubic potential
\bea
\label{eq:pot:inflection}
v\left(\phi\right) = v_0\left[1+\beta\lp\frac{\phi-\phi_0}{\Mp}\rp^{3}\right]\,,
\eea
although our conclusions can be easily generalised to other odd powers. Without loss of generality, we assume $\beta>0$, so the potential is positive when $x>-\beta^{-1/3}$, where we have defined 
\bea
x\equiv \frac{\phi-\phi_0}{\Mp}\, .
\eea
The inflaton is assumed to evolve in the field range comprised between $\phi_\uend$, where inflation ends, and an upper bound value $\phiuv$.

The potential being exactly flat around $\phi_0$, stochastic effects dominate in the neighbourhood of the inflection point. More precisely, in \Refs{Vennin:2015hra, Pattison:2017mbe}, it is shown that, in general, when the condition
\bea \label{eq:classicality_criterion}
\frac{v''\,v^2}{v'^2}\gg 1
\eea 
holds, the potential can be assumed to be exactly flat (hence dominated by quantum diffusion), while when the opposite condition applies, the dynamics of the inflaton is essentially classical. The condition~\eqref{eq:classicality_criterion} is saturated at a value of $x$ such that $\vert x \vert = [2 v_0/(3\beta)]^{1/3}\equiv \dphiwell/(2\Mp)$, which defines the width
\bea \label{eq:dphiwell_cubic}
\frac{\dphiwell}{\Mp} = 2 \lp\frac{2}{3}\frac{v_0}{\beta}\rp^{1/3}
\eea 
 of the field range dominated by stochastic diffusion. In other words, the inflection point potential can be approximated as being exactly flat over a field interval (that we call the ``quantum well'') centred at the inflection point and of width given by $\dphiwell$, and outside this range, as giving rise to purely classical dynamics. The situation is depicted in \Fig{fig:inflection_linear_potential}.

As a consequence, if $\phiuv$ is set far outside the quantum well, which is what we do in practice, it does not affect the PDF of the number of \efolds~and it becomes an irrelevant parameter. If $\phiend$ lies outside the quantum well, a constant, deterministic number of \efolds~is realised between the exit of the quantum well and $\phiend$, so it only shifts the PDF of the number of \efolds~by a constant value. 

\paragraph{Slow-roll conditions}
A few words are in order regarding the slow-roll conditions. Since $\beta x^3 \propto v_0$ at the boundaries of the quantum well, and given that $v_0$, which measures the potential energy in Planckian units, must be much smaller than $1$, the potential is almost constant over the quantum well, and the potential slow roll parameters in that region are controlled by $v_{x}/v \simeq 3\beta x^2$, and $v_{xx}/v\simeq 6\beta x$. Computing these two quantities at the edges of the quantum well, one finds $\beta^{1/3} v_0^{2/3}$ and $\beta^{2/3} v_0^{1/3}$ respectively, so the quantum well is within the slow-roll regime as long as
\bea
\label{eq:flat:inflection:point:SR:condition}
\beta\ll \frac{1}{\sqrt{v_0}}\, .
\eea
This does not guarantee that the full field range comprised between $\phi_\uend$ and $\phiuv$ is within the slow roll regime, but given that, as soon as $\phi_\uend$ and $\phiuv$ are outside the quantum well, they do not (or only trivially) affect the PDF we are aiming to compute, the condition~\eqref{eq:flat:inflection:point:SR:condition} is sufficient in practice. 

Let us also mention that, if $\beta\ll 1$, then the potential slow-roll conditions are always satisfied above the inflection point, \ie for all $x\geq 0$. If this is not the case however, \ie if $1\ll \beta \ll 1/\sqrt{v_0}$, slow roll is strongly violated around $x\sim \beta^{-1/3}$, so starting from an initial large-field value, one could enter the quantum well away from the slow-roll attractor, even though a slow-roll solution exists there~\cite{Pattison:2018bct}. This is why, here, we view \Eq{eq:pot:inflection} only as an expansion of the potential around the flat inflection point, and assume that, at large-field values, the potential is modified such that one always approaches the quantum well with initial conditions located on the slow-roll attractor. In any case, as explained in \Sec{sec:tail_curvature}, stochastic inflation can be formulated in full phase space, which could allow one to study setups where slow roll is explicitly violated.
\paragraph{Flat quantum-well approximation}
The PDF of the number of \efolds~realised in the potential~(\ref{eq:pot:inflection}) can be computed numerically, by solving \Eq{eq:diff:chi} for the characteristic function and Fourier transforming the result along \Eq{eq:pdf:chi}. Below, we will compare this result with the approximation outlined above, where the dynamics is classical outside the quantum well, and undergoes pure quantum diffusion inside the well. In this approximation, starting from a certain initial field value $\phi$ inside the quantum well, one can write the realised number of \efolds~as
\bea
\N = \N_\mathrm{well} \left(\phi\right) + N_\ucl\left(\phi_0-\dphiwell/2 \to \phiend\right),
\eea 
where the PDF of $\N_\mathrm{well} (\phi)$ has been computed in \Sec{sec:flat_potential}, and $N_\ucl\left(\phi_0-\dphiwell/2 \to \phiend\right)$, which we will simply denote $N_\ucl$ in what follows, stands for the classical, deterministic number of \efolds~realised between the exit of the quantum well, at $\phi=\phi_0-\dphiwell/2$, and the end of inflation. Therefore, the PDF of $\N$ is given by
\bea
P_\phi\left(\N\right) = P_\phi^\mathrm{flat}\left(\N- N_\ucl\right), 
\eea
where $P_\phi^\mathrm{flat}$ is given by \Eq{eq:flat_PDF} with $\dphiwell$ given by \Eq{eq:dphiwell_cubic}. If one starts from an initial value of $\phi$ located beyond the quantum well, \ie $\phi>\phi_0+\dphiwell/2$, then one simply has to add another classical contribution to the total number of \efolds, \ie 
\bea
\N = N_\ucl\left(\phi \to \phi_0+\dphiwell/2\right) + \N_\mathrm{well} \left(\phi_0+\dphiwell/2\right) + N_\ucl\left(\phi_0-\dphiwell/2 \to \phiend\right),
\eea
which we simply write as $\N = N_\ucl + \N_\mathrm{well} (\phi_0+\dphiwell/2)$, and this gives rise to
\bea
P_\phi\left(\N\right) = P_{\phi_0+\dphiwell/2}^\mathrm{flat}\left[\N- N_\ucl\left(\phi\right)\right] .
\eea

Shifting the PDF by a constant number of \efolds~does not change its decay rates, so the eigenvalues $\Lambda_n$ are given by \Eq{eq:flat_poles}, and making use of \Eqs{eq:mu} and~\eqref{eq:dphiwell_cubic}, one obtains
\bea \label{eq:inflection_poles}
\Lambda_n=
\left(\frac{9v_0\beta^2}{4}\right)^{1/3} \frac{\pi^2}{4} \left(n+\frac{1}{2}\right)^2.
\eea
Let us stress that, because of the slow-roll condition~\eqref{eq:flat:inflection:point:SR:condition}, $v_0\beta^2\ll 1$, the first eigenvalues are necessarily small, which means that the tails are very much unsuppressed in this model. This will have strong consequences for PBH formation, that we will discuss in \Sec{sec:pbh}.  The way that the coefficients $a_n$ in the expansion~\eqref{eq:pdf_inegral_contour} change under a constant shift in the number of \efolds~is also trivial to establish, and this leads to 
\bea
\label{eq:flat:inflection:an:appr}
a_n\left(\phi\right) &= a_n^\mathrm{flat}\left(\phi\right) \ee^{\Lambda_n N_\ucl}\\
&= (-1)^n \frac{\pi}{4}\left(\frac{9}{4}v_0\beta^2\right)^{1/3}\left(2n+1\right) \cos\left[\frac{\pi}{2}\left(2n+1\right)\left(\frac{\phi-\phi_0}{\dphiwell}-\frac{1}{2}\right)\right] \ee^{\Lambda_n N_\ucl},
\eea
where we have made use of \Eq{eq:an:flat} to evaluate $a_n^\mathrm{flat}$. This gives rise to
\bea
\label{eq:PDF:flat:inflection:point}
P_\phi\left(\N\right) = &\frac{\pi}{4}\left(\frac{9}{4}v_0\beta^2\right)^{1/3}  \sum_n (-1)^n \left(2n+1\right) 
\\ & \times
\cos\left[\frac{\pi}{2}\left(2n+1\right)\left(\frac{\phi-\phi_0}{\dphiwell}-\frac{1}{2}\right)\right] \ee^{- \left(\frac{9v_0\beta^2}{4}\right)^{1/3} \frac{\pi^2}{4} \left(n+\frac{1}{2}\right)^2 \left(\N-N_\ucl\right)}\, ,
\eea
which can be rewritten in terms of the first elliptic theta function,
\bea
P_\phi\left(\N\right) = &\frac{\pi}{8}\left(\frac{9}{4}v_0\beta^2\right)^{1/3} 
\vartheta_1^\prime\left[\frac{\pi}{2}\left(\frac{\phi-\phi_0}{\dphiwell}-\frac{1}{2}\right),\ee^{- \left(\frac{9v_0\beta^2}{4}\right)^{1/3} \frac{\pi^2}{4}  \left(\N-N_\ucl\right)}\right]\, .
\eea
The above expressions are derived assuming that one starts from inside the well, otherwise, $\phi$ has to be replaced with $\phi_0+\dphiwell/2$ if $\phi>\phi_0+\dphiwell/2$. 

\begin{figure}[t]
\centering 
\includegraphics[width=.49\textwidth]{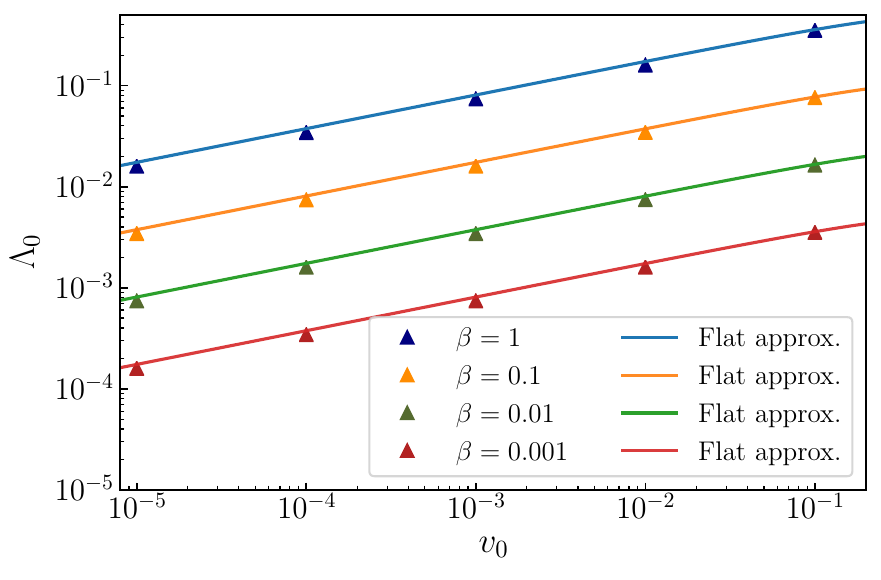}
\includegraphics[width=.49\textwidth]{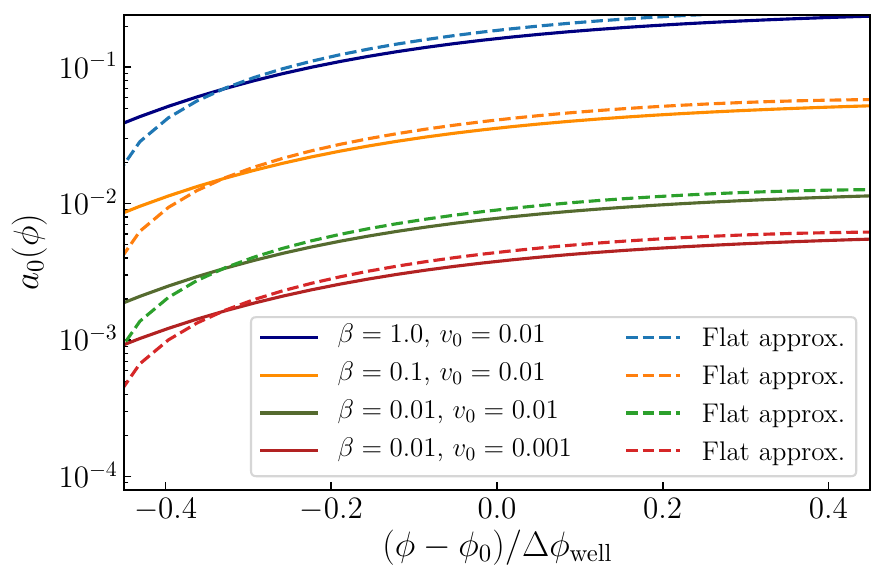}
 \caption{Tail of the PDF of the number of \efolds, $P_\phi(\N)\simeq a_0(\phi) \ee^{-\Lambda_0 \N}$, for the flat inflection-point potential~(\ref{eq:pot:inflection}). The left panel displays the decay rate $\Lambda_0$ as a function of $v_0$ and for a few values of $\beta$. The symbols stand for the full numerical results, while the solid lines stand for the flat quantum-well approximation, \Eq{eq:inflection_poles}.  The right panel shows the amplitude $a_0(\phi)$ for different $v_0$ and  $\beta$. There, the solid lines stand for the full numerical result, and the dashed lines for the flat quantum-well approximation, \Eq{eq:flat:inflection:an:appr}.  In order to satisfy the slow-roll conditions we choose $(\phiuv-\phi_0)/\Mp=1/\beta^{1/3}$, $\phiend=0$ and $\phi_0/\Mp=\dphiwell/2 +0.3/\sqrt{\beta}$.
}
 \label{fig:L0v0_cubic}
\end{figure}
These expressions are compared with a full numerical result in \Fig{fig:L0v0_cubic}. One can see that the leading decay rate, $\Lambda_0$, and the behaviour of $a_0(\phi)$ close to the flat inflection point at $\phi\sim\phi_0$, are accurately reproduced by our approximations~\eqref{eq:inflection_poles} and~\eqref{eq:flat:inflection:an:appr}. At the edges of the quantum well, \ie when $\phi-\phi_0 = \pm \dphiwell/2$, the approximation for $a_0$ starts to deviate from the numerical result, as expected. This otherwise confirms the validity of the approach presented here. Notice that since the expression for $\dphiwell$ comes from saturating the condition~\eqref{eq:classicality_criterion}, $\dphiwell$ in \Eq{eq:dphiwell_cubic} is only defined up to an overall constant of order one, hence so is the case of $\Lambda_n$ in \Eq{eq:inflection_poles} and of $a_n$ in \Eq{eq:flat:inflection:an:appr}. 

\begin{figure}[t]
\centering 
\includegraphics[width=.59\textwidth]{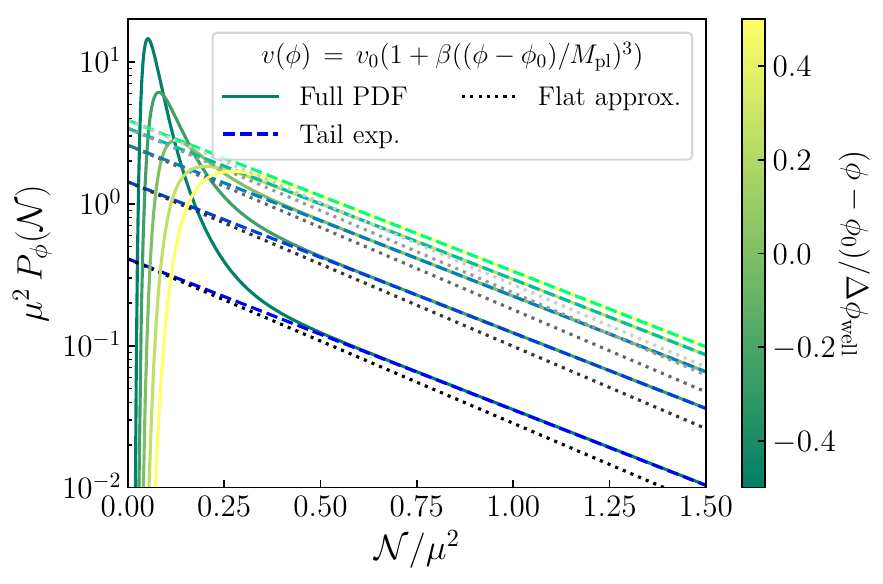}
 \caption{Probability distribution function of the number of \efolds~$\N$ realised in the flat inflection-point potential~(\ref{eq:pot:inflection}), starting from different initial field values $\phi$ labeled in the colour bar. The solid lines stand for the full PDF, while the dashed lines correspond to the leading term in the tail expansion, $P_\phi(\N)\simeq a_0(\phi) \ee^{-\Lambda_0 \N}$. Both are obtained from numerically solving \Eq{eq:diff:chi}. On the contrary, the dotted lines represent the leading term in the flat approximation using the analytical expression for $a_0(\phi)$ and $\Lambda_0$ given in \Eqs{eq:flat:inflection:an:appr} and \eqref{eq:inflection_poles} respectively. To make the comparison with \Fig{fig:PDF_flat} easy, the number of \efolds~is rescaled by $\mu^2=\dphiwell^2/(v_0 \Mp^2)$, where $\dphiwell$ is the width of the region where quantum diffusion dominates, and is given by \Eq{eq:dphiwell_cubic}.
 In order to satisfy the slow-roll condition (\ref{eq:flat:inflection:point:SR:condition}), we have chosen $v_0=0.01$, $\beta=0.01$, $(\phiuv-\phi_0)/\Mp=1/\beta^{1/3}$, $\phiend=0$ and $\phi_0/\Mp=\dphiwell/2 +0.3/\sqrt{\beta}$. 
 One can see that the analytic, flat approximation accurately predicts the amplitude of the tail with a small deviation in the slope.
}
 \label{fig:PDF_cubic}
\end{figure}

In \Fig{fig:PDF_cubic}, we also display the full PDF, computed numerically from solving \Eq{eq:diff:chi} for the characteristic function and Fourier transforming the result along \Eq{eq:pdf:chi}. The result is compared with the leading-tail expansion $P_\phi(\N)\simeq a_0(\phi) \ee^{-\Lambda_0 \N}$, where $\Lambda_0$ and $a_0$ are obtained numerically from searching for the first pole of the solution to \Eq{eq:diff:chi}. Let us note that, when doing so, the fact that \Eq{eq:inflection_poles} provides a good approximation to the pole location $\Lambda_0$ turns out to be very convenient, since it sets an initial value around which to look for the pole, which greatly simplifies the computational problem. One can check that the leading-tail expansion provides an excellent approximation to the full PDF on its tail, as expected. 
On the other hand, the dotted lines, representing the flat approximation given by $a_0(\phi)$ and $\Lambda_0$ in \Eqs{eq:flat:inflection:an:appr} and \eqref{eq:inflection_poles} respectively, show that these simple, analytical formulas provide the right order of magnitude for the amplitude and decay rate of the tail. 

\subsection{Tilted inflection-point potentials}
\label{sec:inflection_linear_potential}

We now consider the possibility that the inflection point is not exactly flat, \ie $v''=0$ at $\phi=\phi_0$ but $v'\neq 0$. Tilted-inflection point potentials of this class~\cite{Garcia-Bellido:2017mdw, Ezquiaga:2017fvi} can be constructed by adding a linear slope to our previous cubic potential~\eqref{eq:pot:inflection}, \ie 
\bea
\label{eq:pot:inflection_linear}
v\left(\phi\right) = v_0\left[1+\alpha\lp\frac{\phi-\phi_0}{\Mp}\rp+\beta\lp\frac{\phi-\phi_0}{\Mp}\rp^{3}\right]\, ,
\eea
where we assume $\alpha\geq 0$ and $\beta\geq 0$. The region of the potential where quantum diffusion dominates still has to be determined from the criterion~\eqref{eq:classicality_criterion}. The quantity $v'' v^2/(v')^2$ vanishes at the inflection point $x=0$. Around this point, if $\alpha$ and $\beta$ are small, there is always a slow-roll region where $v\simeq v_0$. In this regime, $v'' v^2/(v')^2$ is maximal at $x=\pm \sqrt{\alpha/\beta}/3$, where its value is $9v_0\sqrt{\beta}/(8 \alpha^{3/2})$. Two cases need therefore to be distinguished, depending on whether this quantity is smaller or larger than one.
\subsubsection{A single constant-slope well}
In the case where
\bea
\label{eq:quasi:flat:inflection:point:cond:alpha:nowell}
\alpha\gg (v_0^2 \beta)^{1/3},
\eea 
there is no region where the potential can be approximated as quasi constant, since $v'' v^2/(v')^2$ is never larger than one. So there is no almost-constant quantum well of the kind studied in \Sec{sec:flat_potential}. When $\vert x \vert \ll \sqrt{ \alpha/(3\beta) }$ however, the potential slope is almost constant, so the results derived in \Sec{sec:linear_potential} can be applied, over a field range of width 
\bea
\label{eq:dphiwell:tilted:inflection:point}
\dphiwell \simeq 2 \Mp \sqrt{\frac{ \alpha}{3\beta}}.
\eea 
Let us note that this well falls far within the almost constant regime, where both $\alpha x$ and $\beta x^3$ are much less than one, if $\alpha\ll \beta^{1/3}$, and the potential slow-roll conditions, $\Mp v'/v \ll 1$ and $\Mp^2v''/v\ll 1$, reduce to
\bea
\label{eq:cond:sr:tilted:inflection:case:constant:slope}
\alpha\ll 1\, ,\quad\quad \alpha\beta\ll 1\, ,
\eea
 inside the wells. 

\begin{figure}[t]
\centering 
\includegraphics[width=.49\textwidth]{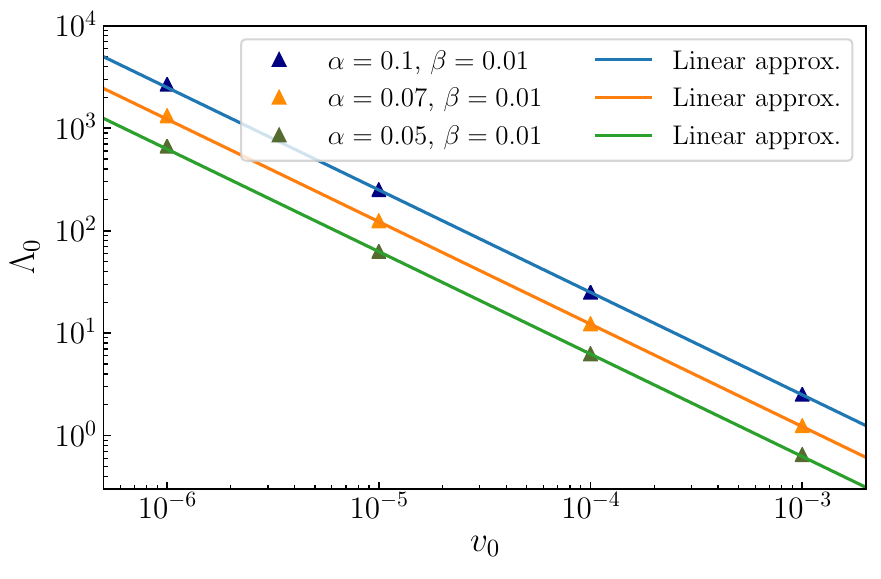}
\includegraphics[width=.48\textwidth]{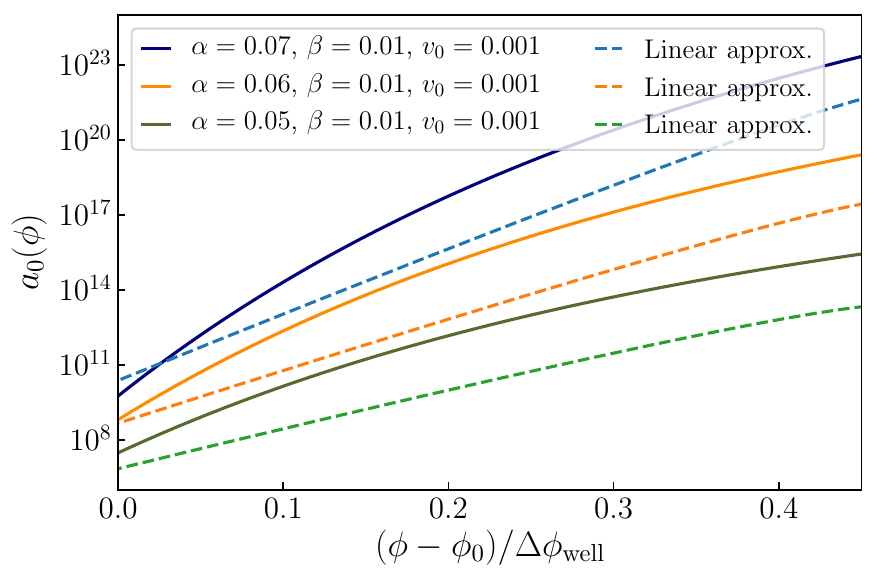}
 \caption{Tail of the PDF of the number of \efolds, $P_\phi(\N)\simeq a_0(\phi) \ee^{-\Lambda_0 \N}$, for the tilted inflection-point potential~(\ref{eq:pot:inflection_linear}). The left panel displays the decay rate $\Lambda_0$ as a function of $v_0$ for a few values of $\alpha$ and $\beta$. The symbols stand for the full numerical results, while the solid lines stand for the linear wide-well approximation, \Eq{eq:quasi_poles}.  The right panel shows the amplitude $a_0(\phi)$ for different $v_0$, $\alpha$ and  $\beta$. There, the solid lines stand for the full numerical result, and the dashed lines for the linear wide-well approximation, \Eq{eq:an:inflection:wide:constant:well}.  In order to satisfy the slow-roll conditions we choose $(\phiuv-\phi_0)/\Mp=\phi_0/\Mp=0.1/\alpha$ and $\phiend=0$.
}
 \label{fig:L0v0_quasi}
\end{figure}

The relation~\eqref{eq:dphiwell:tilted:inflection:point} gives rise to $\dphiwell\alpha/(\Mp v_0) = 2 \alpha^{3/2}/(3 \beta v_0)^{1/2}$, which is much larger than one because of \Eq{eq:quasi:flat:inflection:point:cond:alpha:nowell}. This means that the condition~\eqref{eq:constant:slope:expansion:transcendental:condition} is satisfied, hence we are in the wide-well regime. This implies that \Eq{eq:lambda_almost_const} applies, namely
\bea \label{eq:quasi_poles}
\Lambda_n \simeq \frac{\alpha^2}{4v_0}+\frac{\pi^2 v_0\Mp^2}{\dphiwell^2}\left(n+1\right)^2\, ,
\eea
together with \Eq{eq:a:constant:slope}, namely
\begin{align}
a_n(\phi) \simeq -(-1)^n \frac{\pi v_0 \Mp^2 }{\dphiwell^2 }2\left(n+1\right) \ee^{\frac{\alpha\dphiwell}{2 v_0\Mp} \left(\frac{\phi-\phi_0}{\dphiwell}+\frac{1}{2}\right)} \sin\left[{\pi}\left(n+1\right)\left(\frac{\phi-\phi_0}{\dphiwell}-\frac{1}{2}\right)\right] ,
\label{eq:an:inflection:wide:constant:well}
\end{align}
with a possible additional correction $\ee^{\Lambda_n N_\ucl}$ if a classical number of \efolds~is realised before of after the well, as in \Eq{eq:flat:inflection:an:appr}. Combined together, \Eqs{eq:quasi_poles} and~\eqref{eq:an:inflection:wide:constant:well} lead to the PDF
\bea
P_\phi(\N) &= - \frac{\pi v_0 \Mp^2 }{2 \dphiwell^2 } \ee^{\frac{\alpha\dphiwell}{2 v_0\Mp} \left(\frac{\phi-\phi_0}{\dphiwell}+\frac{1}{2}\right)}
\ee^{-\frac{\alpha^2}{4v_0} \left(\N-N_\ucl\right)} 
 \\ & \quad\quad\quad \times
{\vartheta_4}^\prime \left[\frac{\pi}{2}\left(\frac{\phi-\phi_0}{\dphiwell}-\frac{1}{2}\right),\ee^{-\frac{\pi^2 v_0\Mp^2}{\dphiwell^2}\left(\N-N_\ucl\right)}\right]\, .
\eea

The above approximated formulas for $\Lambda_0$ and $a_0$ are compared with a full numerical solution in \Fig{fig:L0v0_quasi}. On the left panel, one can see that $\Lambda_0$ is accurately reproduced, while on the right panel, only the generic trend and order of magnitude of $a_0$ are accounted for. 
This is because, as already mentioned, the effective values of $\dphiwell$, derived from the saturation of the non-classicality criterion~\eqref{eq:classicality_criterion}, provide estimates up to factors of order one only. The way they enter the PDF for a flat inflection point, \ie the way $\dphiwell$ appears in \Eq{eq:PDF:flat:inflection:point}, is such that this uncertainty produces order-of-one errors in the amplitude of the PDF. However, for a tilted inflection-point in the regime of \Eq{eq:quasi:flat:inflection:point:cond:alpha:nowell}, $\dphiwell$ enters exponentially in the amplitude of the PDF, see \Eq{eq:an:inflection:wide:constant:well}. This implies that these order-one corrections are exponentiated, potentially leading to more substantial corrections in the amplitude of the tail. Let us however stress that our determination of $\Lambda_n$ does not suffer from this issue, and that, as mentioned above, since it provides a first guess for the location of the pole, it plays a crucial role in the numerical determination of the poles and of their residues.

\subsubsection{From two quantum wells separated by a constant slope, to a single quantum well}
If the condition
\bea
\label{eq:quasi:flat:inflection:point:cond:alpha}
\alpha\ll (v_0^2 \beta)^{1/3}
\eea
is realised, there exist two regions where $v'' v^2/(v')^2$ is larger than one, namely for $x\in[-x_+,-x_-]$ and $x\in [x_-,x_+]$, where 
\bea
x_- &\simeq \frac{\alpha^2}{6\beta v_0} \left(1+\frac{\alpha^3}{6\beta v_0^2}+\cdots\right),\\
x_+ &\simeq \frac{\dphiwell}{2}\left[1-\left(\frac{2\alpha^3}{3^4\beta v_0^2}\right)^{1/3}+\cdots\right],
\eea
with $\dphiwell$ given in \Eq{eq:dphiwell_cubic} and where ``$\cdots$'' denotes higher powers of $\alpha^3/(\beta v_0^2)$.  In the limit of \Eq{eq:quasi:flat:inflection:point:cond:alpha}, one has $x_- \ll x_+$, so the two wells are almost adjacent. By computing the relative importance of the terms $\beta x^3$ and $\alpha x$ at the point $\pm x_-$, one notices that it is proportional to $\alpha^3/(\beta v_0^2)$, hence it is very small because of \Eq{eq:quasi:flat:inflection:point:cond:alpha}. Therefore, in the interval $[x_-,x_+]$, the potential is of the quasi constant-slope type. One has therefore three wells in series: a first quasi-constant well between $x_+$ and $x_-$, a quasi constant-slope well between $x_-$ and $-x_-$, and a second quasi-constant well between $-x_-$ and $-x_+$. 

Let us note that, if \Eq{eq:quasi:flat:inflection:point:cond:alpha} is satisfied, these wells are far within the almost constant regime where both $\alpha x$ and $\beta x^3$ are much less than one, and the potential slow-roll conditions, $\Mp v'/v \ll 1$ and $\Mp^2v''/v\ll 1$, reduce to 
\bea
\label{eq:cond:alpha:lt:1}
\alpha\ll 1
\eea
and to \Eq{eq:flat:inflection:point:SR:condition} inside the wells. 

The width of the constant-slope well is given by $\dphiwell/\Mp=2 x_- = \alpha^2/(3 \beta v_0)$. One therefore has $\dphiwell \alpha/(\Mp v_0) = \alpha^4/(3\beta v_0^2)$, which is much smaller than one because of \Eqs{eq:quasi:flat:inflection:point:cond:alpha} and~\eqref{eq:cond:alpha:lt:1}. As a consequence, the constant-slope well is in the narrow-well regime, in the sense of \Eq{eq:small:alpha:limit:def}. According to the considerations of \Sec{sec:linear_potential}, this means that we are in fact in the presence of a quasi-constant potential, so the three wells in series are in effect a single, almost-constant well, with a width given by $\dphiwell = 2\Mp x_+$, \ie by \Eq{eq:dphiwell_cubic}. One concludes that, in that case, the same results as those derived in \Sec{sec:inflection_potential} apply. 

\begin{figure}[t]
\centering 
\includegraphics[width=.59\textwidth]{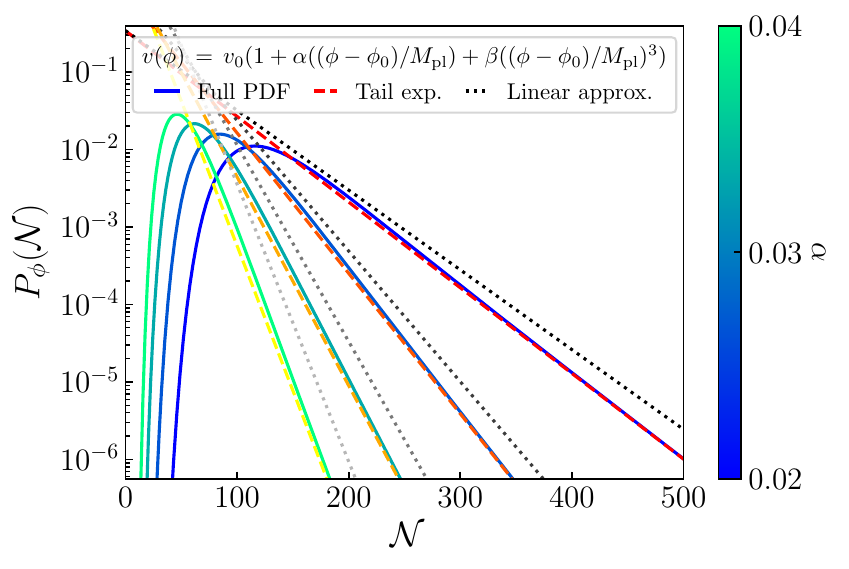}
 \caption{Probability distribution function of the number of \efolds~$\N$ realised in the tilted inflection-point potential~(\ref{eq:pot:inflection_linear}), starting from the inflection point $\phi=\phi_0$, as a function of the linear slope $\alpha$. We compare the full PDF (solid lines) with the leading tail expansion (dashed lines), $P_\phi(\N)\simeq a_0(\phi) \ee^{-\Lambda_0 \N}$, and the linear wide-well approximation given by \Eq{eq:quasi_poles} and \Eq{eq:an:inflection:wide:constant:well}. The $\alpha$-term suppresses the tail of the PDF at large $\N$. 
 We choose $v_0=5\cdot10^{-3}$ and $\beta=10^{-3}$, such that the condition~\eqref{eq:quasi:flat:inflection:point:cond:alpha:nowell} applies, and $(\phiuv-\phi_0)/\Mp=\phi_0/\Mp=0.1/\alpha$ and $\phiend=0$.
 }
 \label{fig:PDF_inflection_linear}
\end{figure}

In summary, when a tilt is introduced into  a flat inflection-point model, as long as the slope $\alpha$ is smaller than the bound~\eqref{eq:quasi:flat:inflection:point:cond:alpha}, it has no effect. When it is larger, it changes the almost constant well into an almost constant-slope well, and adds a contribution $\alpha^2/(4 v_0)$ to the eigenvalues, hence suppresses the tails. This can be clearly seen in \Fig{fig:PDF_inflection_linear}, where the PDF of the number of \efolds~is shown for various values of the slope $\alpha$ [notice that, as in \Fig{fig:L0v0_quasi}, in the linear wide-well approximation, only the order of magnitude of the amplitude of the tail is correctly reproduced, while its decay rate is accurately accounted for, see the discussion below \Eq{eq:an:inflection:wide:constant:well}]. 
%

\section{Implications for primordial black hole formation}
\label{sec:pbh}
We have seen that quantum diffusion makes the tail of the PDF of the duration of inflation decay exponentially with the number of $e$-folds. We have exemplified this phenomenon with several toy models including flat, linear, flat inflection-point and tilted inflection-point potentials. The non-Gaussian nature of the tail of the PDF introduces important differences with the standard classical picture of quasi-Gaussian distributions, which translates into important differences for the predicted amount of PBHs, that we now discuss. 

\begin{figure}[t!]
\centering 
\includegraphics[width=.49\textwidth]{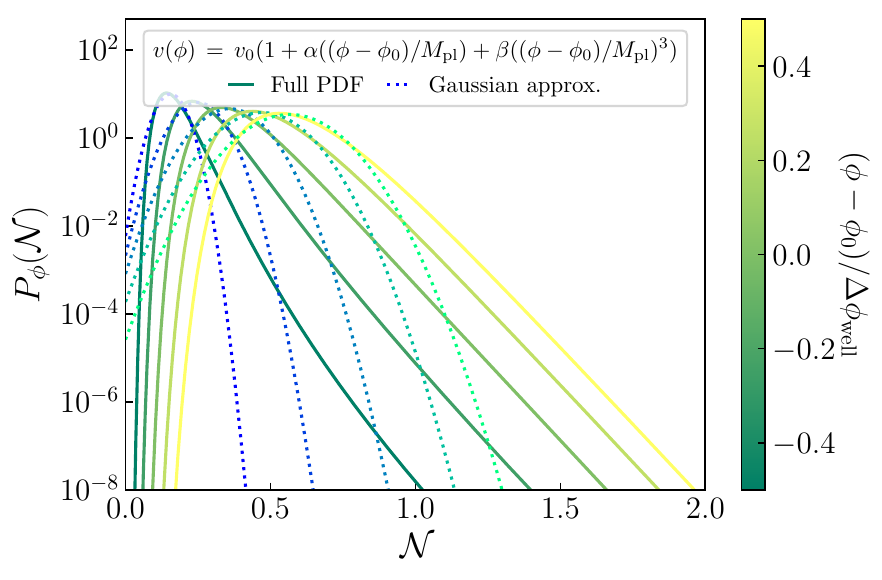}
\includegraphics[width=.49\textwidth]{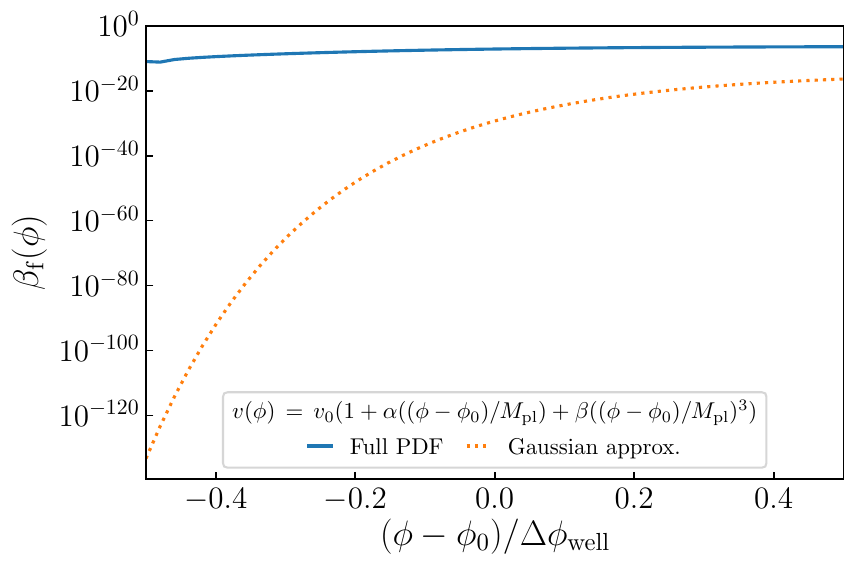}
 \caption{Comparison of the full calculation of the probability density function (PDF) of the curvature perturbations with the Gaussian approximation. On the left panel we present the PDF as a function of the number of e-folds $\mathcal{N}$ for different initial field values $\phi$. The Gaussian approximation does not capture the exponential decay of the tail. As shown in the right panel, this leads to a very significant difference in the abundance of PBHs $\beta_\text{f}(\phi)$. For this example we have chosen a tilted inflection-point potential with $v_0=10^{-3}$, $\alpha=0.24$, $\beta=9$, $\phiuv-\phi_0=\phi_0=0.7\Mp/\alpha$ and $\phiend=0$, consistent with the slow-roll approximation.}
 \label{fig:full_vs_gaussian}
\end{figure}

In order to relate the coarse-grained curvature perturbation with the number of \efolds, one can use the relation (\ref{eq:cg-deltaN}), where the mean number of \efolds~can be computed directly from the characteristic function by making use of \Eq{eq:mean:N:chi}.\footnote{Alternatively, the mean number of \efolds~can be obtained by solving the differential equation~\cite{Vennin:2015hra}
\bea
\langle\N\rangle''-\frac{v_\phi}{v^2}\langle\N\rangle'+\frac{1}{v\Mp^2}=0\, ,
\eea
with boundary conditions $\langle\N\rangle(\phiend)=\langle\N\rangle'(\phiuv)=0$. This equation follows directly form the definition of the characteristic function \eqref{eq:characteristicFunction:def} and the differential equation \eqref{eq:diff:chi} it satisfies. Combining \Eqs{eq:chi:pole:expansion} and~\eqref{eq:mean:N:chi}, one also has
\bea
\langle \N(\phi) \rangle = \sum_n\frac{a_n(\phi)}{\Lambda_n^2}\, .
\eea}  
By integrating the PDF~\eqref{eq:pdf_inegral_contour} above the threshold $\N_\uc = \langle \N \rangle + \zeta_\uc$, one obtains from \Eq{eq:beta:pdf} the mass fraction of PBHs,
\bea
\label{eq:beta:sum}
\beta_\mathrm{f}(\phi)=\sum_{n}\frac{1}{\Lambda_n}\,a_n(\phi)\,e^{-\Lambda_n\left[\zeta_\uc+\langle\N\rangle(\phi)\right]}\, .
\eea
This should be contrasted with the standard classical result, where \Eq{eq:tail_expansion:classical} gives rise to
\bea
\label{eq:beta:class}
\beta_{\mathrm{f}}^\ucl(\phi)=\frac{\int_{\bar{k}(\phi)}^{k_\uend}\calP_{\zeta,\mathrm{cl}}\dd\ln k}{\sqrt{2\pi}\zeta_\uc}\,\exp\left[-\dfrac{\zeta_\uc^2}{2\int_{\bar{k}(\phi)}^{k_\uend}\calP_{\zeta,\mathrm{cl}}\dd\ln k}\right]\,,
\eea
which depends exponentially on the square of $\zeta_\uc$, rather than on $\zeta_\uc$ directly as in \Eq{eq:beta:sum}, and which leads to estimates of the mass fraction that can be orders of magnitude away from the actual result~\eqref{eq:beta:sum} (see \Fig{fig:full_vs_gaussian} below for a particular example). 
Let us now review the potentials discussed in \Sec{sec:applications}. 
\paragraph{Flat potential}
In a flat potential, in the notations of \Sec{sec:flat_potential}, the mean number of \efolds~is given by $\langle \N \rangle = \mu^2 x(1-x/2)$, which also corresponds to the $\alpha\to 0$ limit of \Eq{eq:mean:N:constant:slope}. Using the formulas derived in \Sec{sec:flat_potential}, \Eq{eq:beta:sum} gives rise to
\bea
\label{eq:beta:zetac:flat:potential}
\beta_\mathrm{f}(\phi)=\sum_n \frac{4}{(2n+1)\pi}\sin\left[\frac{\pi}{2}\left(2n+1\right)x\right]\,\ee^{-\pi^2\left(n+\frac{1}{2}\right)^2\left[\frac{\zeta_\uc}{\mu^2}+  x(1-\frac{x}{2})\right]}\, .
\eea
This expression is always well approximated by its first term, and one can see that for PBHs not to be over produced, one needs to impose
\bea
\label{eq:mu:cond:PBH}
\mu\ll \sqrt{\zeta_\uc}\, ,
\eea
in agreement with the conclusions of \Ref{Pattison:2017mbe}. This places an upper bound on the width, or a lower bound on the height, of flat sections in the potential. 
\paragraph{Constant-slope potential}
In a constant-slope potential, two regimes have to be distinguished. In the narrow-well regime, defined by \Eq{eq:small:alpha:limit:def}, the same results as for the flat potential apply, and one recovers \Eq{eq:mu:cond:PBH}. In the wide-well regime, defined by \Eq{eq:constant:slope:expansion:transcendental:condition}, the mean number of \efolds~is given by \Eq{eq:Ncl:constant:slope}, and using the results of \Sec{sec:linear_potential}, \Eq{eq:beta:sum} gives rise to
\bea
\beta_\mathrm{f}^\mathrm{wide}(\phi)=&
\frac{8 v_0 \pi}{\alpha^2 \mu^2}
\frac{ \ee^{\frac{\alpha}{4 v_0}\left(\frac{\phiuv}{\Mp}x-\alpha\zeta_\uc\right)}}{1+4 \pi^2\frac{ v_0^2  \Mp^2}{\alpha^2\phiuv^2}(n+1)^2}
\\ & \times
\sum_n (-1)^{n+1} (n+1)\sin\left[\pi(n+1)(x-1)\right] \ee^{-\pi^2 (n+1)^2\left(\frac{\zeta_\uc}{\mu^2}+x \frac{v_0\Mp}{\alpha\phiuv}\right)}
\, .
\eea
In the limit of \Eq{eq:constant:slope:expansion:transcendental:condition}, the argument of the overall exponential always dominates over the one of the exponential in the sum (at least for the first few terms), so in order to avoid overproduction of PBHs in this model, one must have
\bea
\label{eq:alpha:min:overproduction}
\alpha \gg  \max\left(\sqrt{v_0},\frac{\phiuv}{\Mp}\right)\, .
\eea

\paragraph{Flat inflection-point potential}
\begin{figure}[t]
\centering 
\includegraphics[width=.49\textwidth]{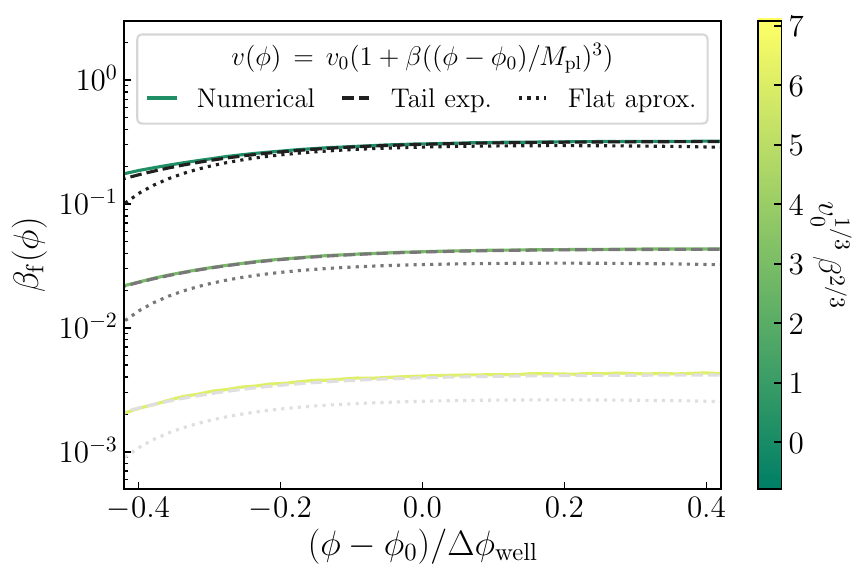}
\includegraphics[width=.49\textwidth]{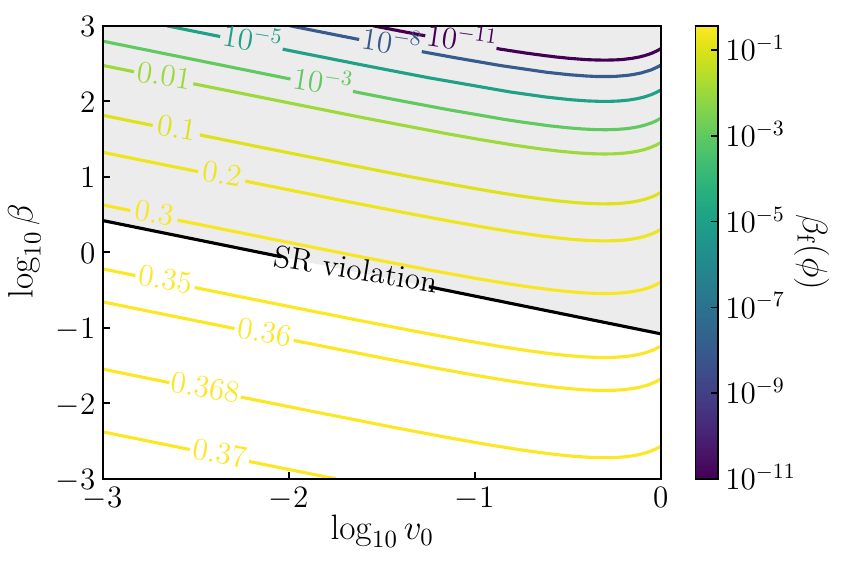}
\caption{PBH abundance $\beta_\mathrm{f}(\phi)$ in a flat inflection-point potential. Left panel: $\beta_\mathrm{f}(\phi)$ is displayed as a function of the initial field value $\phi$ for $\zeta_\uc=1$ and different choices of the combination of parameters $v_0\beta^2$, which controls the tail of the PDF, see \Eq{eq:mu:cubic:inflection:point}. The solid lines stand for a full numerical result, the dashed lines for the numerical result if only the dominant term is kept in the tail expansion (\ref{eq:beta:sum}), and the dotted lines to the constant-well approximation, \ie to \Eq{eq:beta:zetac:flat:potential} with $\mu$ given by \Eq{eq:mu:cubic:inflection:point}. 
For the plot we fix $v_0=10^{-2}$ and vary $\beta$.  
Right panel: Contour plot of $\beta_\text{f}(\phi)$ as a function of the parameters $v_0$ and $\beta$ for $\zeta_c=1$. The grey shaded region corresponds to where the slow-roll approximation does not hold across the entire range $\phi_0-\dphiwell/2$ to $\phi_0+\dphiwell/2$, \ie to where \Eq{eq:flat:inflection:point:SR:condition} is not satisfied. One can see that, when slow roll is satisfied, PBHs are overproduced.
}
 \label{fig:beta_inflection}
\end{figure}
As explained in \Sec{sec:inflection_potential}, a flat inflection-point potential is equivalent to a flat potential with $\dphiwell$ given by \Eq{eq:dphiwell_cubic}, \ie $\mu$ given by 
\bea
\label{eq:mu:cubic:inflection:point}
\mu^2=4  \left(\frac{2}{3\beta\sqrt{v_0}}\right)^{2/3}\, .
\eea
This parameter is necessarily large because of the slow-roll condition~\eqref{eq:flat:inflection:point:SR:condition}, so according to the above considerations, see \Eq{eq:mu:cond:PBH}, $\beta_\mathrm{f}$ is large in this model, which is confirmed by the numerical results displayed in \Fig{fig:beta_inflection}. We therefore reach the interesting conclusion that PBHs are always overproduced in a flat inflection point potential, if one approaches the inflection point along the slow-roll attractor. 

This is also consistent with the results of \Ref{Pattison:2017mbe}, see Sec.~5.4 of that reference, where the same conclusion was reached for potentials of the type $v=v_0(1+\alpha\phi^p)$, although these potentials were restricted to positive field values. This suggests that our findings are independent of the order of the polynomial (here cubic) that realises the flat inflection point.

\begin{figure}[t]
\centering 
\includegraphics[width=.49\textwidth]{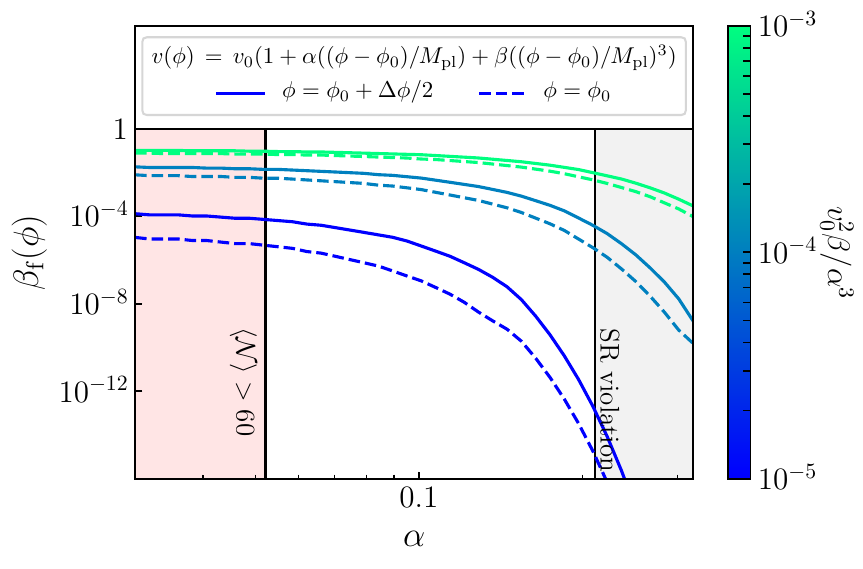}
 \caption{PBH abundance $\beta_\text{f}(\phi)$ for a tilted-inflection point potential as a function of the slope $\alpha$ for $\zeta_c=1$. The combination $v_0^2\beta/\alpha^3$ is shown in the colour bar, and is such that the condition~\eqref{eq:quasi:flat:inflection:point:cond:alpha:nowell} applies (otherwise, the model would be equivalent to a flat inflection-point potential). We set $\beta=50\alpha^3$. For reference, we shade in grey the region where $\alpha\cdot \beta>0.1$ and the slow-roll condition, \Eq{eq:cond:sr:tilted:inflection:case:constant:slope}, is violated.  
 Then, for this example the mean number \efolds{} to cross the inflection point is given by $\langle\N\rangle\sim 1/6\alpha^2$. We shade in red the region where the crossing time is larger than 60 \efolds{}. The bounds $\phiend$ and $\phiuv$ are set according to $\phiuv-\phi_0=\phi_0=0.1\Mp/\alpha$ and $\phiend=0$. We also evaluate $\beta_\text{f}(\phi)$ at different initial field values $\phi$ indicated by the solid/dashed lines. For this range of $v_0^2\beta/\alpha^3$ we thus verify numerically that one either overproduces PBHs or violates the slow-roll condition.
}
 \label{fig:beta_inflection_linear}
\end{figure}

\paragraph{Tilted inflection-point potential} If the inflection-point potential is tilted with a slope $\alpha$ smaller than the upper bound~\eqref{eq:quasi:flat:inflection:point:cond:alpha}, the effect of the slope is negligible and one recovers a quasi flat inflection-point potential, which we just saw overproduces PBHs. If $\alpha$ is larger however, such that the condition~\eqref{eq:quasi:flat:inflection:point:cond:alpha:nowell} is satisfied, one recovers an almost constant-slope potential in the wide-well regime, with $\dphiwell$ given by \Eq{eq:dphiwell:tilted:inflection:point}. This implies that $\alpha\Mp/\dphiwell \propto \sqrt{\alpha\beta} $, which is much smaller than one because of \Eq{eq:cond:sr:tilted:inflection:case:constant:slope}. Therefore, the second of the conditions~\eqref{eq:alpha:min:overproduction} is not satisfied, and PBHs are overproduced too. This is confirmed by the numerical results of \Fig{fig:beta_inflection_linear} where one can check that, when $\beta = 50 \alpha^3$ and $v_0^2\beta/\alpha^3>10^{-5}$, $\beta_\text{f}(\phi)\gtrsim10^{-16}$  as soon as the slow-roll conditions are satisfied [which here are taken to be  $\alpha\cdot\beta=0.1$, see \Eq{eq:cond:sr:tilted:inflection:case:constant:slope}]. We also highlight in red the region where the mean number of \efolds{}  taken to cross the inflection point is larger than $60$, which is inconsistent with CMB observations. 
It is to be noted that as $v_0^2\beta/\alpha^3$ is reduced, the distance between the poles of the characteristic function decreases, and one approaches a continuum of eigenvalues $\Lambda_n$. Higher-order terms in the expansion~\eqref{eq:tail_expansion} then become important, and we find the PBH abundance to be smaller than what the leading term alone would suggest. As a consequence, although it remains true that in general, PBHs are overproduced if one approaches the inflection point without violating slow roll, it may be possible to design fine-tuned situations where slow roll is only mildly violated and the overproduction problem is avoided. 
In addition, we compare in \Fig{fig:full_vs_gaussian} the PBH abundance obtained by solving the full PDF, with the Gaussian approximation. Because the latter does not capture the exponential tail, it underestimates $\beta_\mathrm{f}$ by many orders of magnitude.

\section{Discussion}
\label{sec:discussion}

Primordial curvature perturbations set the initial conditions for the standard model of cosmology. At large scales they determine the seeds of the cosmic structures we observe in the anisotropies of the CMB. This gives us precise measurements of the amplitude and tilt of the primordial power spectrum, as well as constraints on the leading-order non-Gaussian corrections, in the $\sim 7$ \efolds~range of inflation where the scales observed in the CMB cross out the Hubble radius. At small scales, however, non-linear structures in the universe prevent us from probing the nature of the primordial fluctuations. This leaves most of the inflationary potential after the CMB scales are generated unconstrained.  

One possible tracer of the late-time inflationary evolution is PBHs. If the curvature perturbations are sufficiently large, they collapse upon horizon reentry during the radiation-dominated era. Their mass is associated with the horizon size at the time of reentry, which can be linked to the time when the curvature perturbations are generated during inflation. Therefore, by measuring the abundance of PBHs and their mass spectrum one could reconstruct the behavior of the primordial curvature perturbations at scales much smaller than those probed by the CMB.

Since PBHs form from rare, large curvature perturbations, in this work we have explored how to properly compute the tail of their distributions. We have developed a novel approach to compute these tails, from the poles and residues of the characteristic function that is derived in the stochastic-$\delta N$ formalism. Contrary to the standard, classical calculation, in which the tails decay in a Gaussian way (with possible polynomial modulations, if $\fnl$ or $\gnl$-type corrections are included), we have found that these tails have in fact an exponential decay. 

Let us mention that some inflationary models are already known to display a $\chi$-square statistics for the curvature perturbation, such as models of axion inflation in which the gauge field sources the curvature perturbations~\cite{Garcia-Bellido:2016dkw}, and that the implications of a $\chi$-square distribution for the gravitational wave signature have been studied in detail in \Ref{Garcia-Bellido:2017aan}. However, here, the presence of exponential tails has been found even in single-field, slow-roll inflationary scenarios. For such models, we have found simple analytical approximations that capture the behaviour of the tails of the PDF, as well as developed efficient numerical techniques to compute them precisely. This allowed us to properly estimate the abundance of PBHs associated with each model. 
Note that because of the exponential decay of the tail, the difference with the Gaussian approximation can be of many orders of magnitude as we exemplify in \Fig{fig:full_vs_gaussian}. 

We have found that potentials featuring regions where quantum diffusion dominates over the classical roll of the field can be either approximated by locally constant potentials, or by locally constant-slope potentials. In the first case, the requirement that PBHs are not overproduced places an upper bound on the squared width of the flat region, divided by its height, see \Eq{eq:mu:cond:PBH}. In the second case, both the width and the height are bounded from above, see \Eq{eq:alpha:min:overproduction}. When applied to inflection-point potentials, regardless of whether the inflection point is flat or tilted, these conditions cannot be satisfied unless slow roll is violated when approaching the inflection point. This is therefore one of the main results of our work: inflection-point models that do not feature slow-roll violations overproduce PBHs. 

Natural extensions of this work would therefore be to consider potentials in which the field trajectories leave the inflationary slow-roll attractor. As explained in \Sec{sec:tail_curvature}, the stochastic-$\delta N$ formalism can be formulated in full phase space~\cite{Grain:2017dqa, Pattison:2019hef}, in which the pole and residue approach presented here still applies. Similarly, multiple-field scenarios could be analysed with the same techniques.

Finally, it is important to emphasise the universality of our results. The tail of the distribution function of the coarse-grained curvature perturbation $\zeta_{\text{cg}}$ induces non-Gaussian deviations, in the form of exponential tails, on {\em all scales}, with amplitude $a_n(\phi)$ and exponential decay $\Lambda_n$. This is because, in a given inflationary model, $\Lambda_n$ is fixed (it does not depend on $\phi$) while $a_n(\phi)$ depends on the scale at which $\zeta_{\text{cg}}$ exits the Hubble radius. Therefore, there is an exponential tail across the whole spectrum of modes, with the same decay rate, although its amplitude depends on the specific inflationary dynamics associated with each scale. 
For plateau-like potentials for instance, these non-Gaussian effects may be significantly relevant, as in the case of quasi-inflection point models for PBH production.

Even if on large scales (probed in the CMB and in the large-scale structures), the effect of exponential tails may be negligible in most models (although this remains to be checked explicitly), on intermediate scales, corresponding to small halos, \eg Lyman-alpha scales, or even smaller, like ultra-compact mini halos, the exponential tail effects may become very relevant. In particular, they could induce an enhancement of the non-linear collapse of structures on small scales that could have important consequences for large-scale structure formation, and thus for interpreting data from future surveys like DESI, Euclid and LSST.

\acknowledgments
We are thankful to Sebastien Clesse, Wayne Hu, Sam Passaglia, Chris Pattison and David Wands for their comments on the draft. 
We would also like to thank Nahid Ahmadi, Zahra Ahmadi, Niloufar Feyz, Mahdiyar Noorbala and Borna Salehian for their insifghtful comments on \Sec{sec:pole:chi}, see footnotes~\ref{footnote:ack} and~\ref{footnote:ack2}. 
JME is supported by NASA through the NASA Hubble Fellowship grant HST-HF2-51435.001-A awarded by the Space Telescope Science Institute, which is operated by the Association of Universities for Research in Astronomy, Inc., for NASA, under contract NAS5-26555. He is also supported by the Kavli Institute for Cosmological Physics through an endowment from the Kavli Foundation and its founder Fred Kavli.
VV acknowledges funding from the European Union's Horizon 2020 research and innovation programme under the Marie Sk\l odowska-Curie grant agreement N${}^0$ 750491.
The authors acknowledge support from the Research Project FPA2015-68048-03-3P (MINECO-FEDER) and the Centro de Excelencia Severo Ochoa Program SEV- 2016-0597.

\appendix

\section{Coarse-grained curvature perturbation}
\label{sec:zetacg}

If a large fluctuation of the curvature perturbation appears across a Hubble patch, this patch may collapse and form a black hole. For causality reasons, it is often argued that whether or not this occurs can only depend on the value of the curvature perturbation inside the patch. This is why the coarse-grained curvature perturbation, defined as the mean value of the curvature perturbation over a Hubble patch, of comoving volume $(aH)^3$, is usually considered,
\bea
\zeta_{\mathrm{cg}} (\bm{x}) = (aH)^3 \int\dd\bm{y} \zeta(\bm{y}) W\left(aH\left\vert \bm{y} - \bm{x}\right\vert \right),
\eea
where $W$ is a window function such that $W(x)\simeq 1$ if $x\ll 1$ and $W(x)\simeq 0 $ if $x\gg 1$, and normalised in the sense that $ 4\pi \int_0^\infty x^2 W(x)\dd x=1$, such that after coarse graining, a constant field remains a constant field of the same value. A usual criterion for PBH formation is that when $\zetacg(\bm{x})$ exceeds a certain threshold $\zeta_\uc$, the Hubble patch centred on $\bm{x}$ collapses and forms a black hole.

The Fourier transform of this coarse-grained curvature perturbation is given by
\bea
\label{eq:def:tilde:W}
\zeta_{\mathrm{cg}} (\bm{k}) = 
\zeta(\bm{k})
\underbrace{
4\pi
\left(\frac{aH}{k}\right)^3 \int_0^\infty W\left(\frac{aH}{k} u \right)  \sin(u) u\dd u}_{ \widetilde{W}\left(\frac{k}{aH}\right)}
,
\eea
which defines $\widetilde{W}$, that shares similar properties with $W$. Indeed, when $aH/k \gg 1$, the values of $u$ such that $W\left(\frac{aH}{k} u \right)$ is not close to zero are very small, so one can replace $\sin(u)\simeq u$ in the integral over $u$, and using the normalisation condition stated above, one obtains $\widetilde{W}\left[k/(aH)\right] \simeq 1$ in that limit. In the opposite limit, when $aH/k \ll 1$, since $W$ is roughly $1$ until $u\sim k/(aH)$, the integral over $u$ in \Eq{eq:def:tilde:W} is of order $k/(aH)$, hence $\widetilde{W}\left[k/(aH)\right]\propto (aH/k)^2\ll 1 $. 

The details of $\widetilde{W}$ between these two limits depend on those of $W$. For instance, if $W$ is a Heaviside step function,
\bea
W(x) = \frac{3}{4\pi} \theta(1-x),
\eea
where $\theta(x)=1$ if $x>0$ and $0$ otherwise, and where the pre-factor is set in such a way that the above normalisation condition is satisfied, \Eq{eq:def:tilde:W} gives rise to
\bea
\label{eq:Fourier:transform:Heaviside}
\widetilde{W}\left(\frac{k}{aH}\right) = 3 \left(\frac{a H}{k}\right)^3 \left[ \sin \left(\frac{k}{a H}\right)-\frac{k}{a H}\cos\left(\frac{k}{a H}\right)\right],
\eea
which verifies the two limits given above. 

There is some freedom in the choice of the window function, and different window functions can lead to rather different results for the PBH abundance~\cite{Ando:2018qdb}. In fact, comparing the locally coarse-grained curvature perturbation with a certain threshold is only an approximated procedure. More realistic approaches incorporate the full real-space profile of the density contrast across the inhomogeneity, either numerically or by studying the compaction function~\cite{Musco:2004ak, Polnarev:2006aa, Musco:2018rwt, Kalaja:2019uju}. Although such analyses reveal the existence of a variety of different situations, depending on the details of the density profile, in most cases the scales that contribute most to forming a PBH of mass $M$ are those around the Hubble radius when it contains that mass: much smaller scales average out inside the inhomogoneity, as the calculation above indicates, and much larger scales simply rescale the local amplitude of the background density. For this reason, we consider a coarse-grained field made up of scales ``around'' the Hubble radius only,
\bea
\label{eq:zetacg:def:1}
\zeta_{\mathrm{cg}}(\bm{x}) = \left(2\pi\right)^{-3/2}\int_{k\sim aH}\dd {\bm{k}}\zeta_{\bm{k}} e^{i\bm{k}\cdot\bm{x}}\, .
\eea
In this expression, ``$k\sim aH$'' implies the existence of a window function. For simplicity, we consider
\bea
\label{eq:zetacg:def}
\zeta_{\mathrm{cg}}(\bm{x}) = \left(2\pi\right)^{-3/2}\int_{aH<k<a_\uend H_\uend}\dd {\bm{k}}\zeta_{\bm{k}} e^{i\bm{k}\cdot\bm{x}}\, ,
\eea
\ie a top hat window function in Fourier space, that selects out modes comprised between the Hubble scale at the time of formation and the one at the end of inflation. Since we consider scales that are generated a few \efolds~before the end of inflation, we are integrating over a few \efolds~of scales as we should. Obviously, the details of the window function (its shape and the precise range of scales) are arbitrary, but what makes this choice convenient is that \Eq{eq:zetacg:def} coincides with the coarse-grained fluctuation in the number of \efolds~computed in the stochastic-$\delta N$ formalism, see \Sec{sec:Stochastic:DeltaN}. Needless to say, one could use another choice of window function and define the coarse-grained curvature perturbation differently. This would imply to coarse grain the fields differently in the stochastic inflation formalism. For instance, if a smooth $\widetilde{W}$ function is employed, the stochastic noise contains contributions from different modes, so the same mode contributes to the realisation of the noise at different times, and the noise becomes coloured. While coloured noises can be dealt with in the stochastic inflation formalism (see \eg \Refs{Casini:1998wr, Winitzki:1999ve, Matarrese:2003ye, Liguori:2004fa}), they are technically more challenging, which explains our choice.

Let us finally note that, in practice, the statistics of the number of \efolds~starting from a fixed field configuration is computed in this work. However, since different regions of the universe realise different amounts of expansion between that field configuration and the end of inflation, a certain fixed physical scale emerges at different field configurations in different patches of the universe. Therefore, the statistics we compute is not exactly the one of the curvature perturbation coarse-grained at a fixed physical scale. However, given the uncertainties in the coarse-graining procedure mentioned above, this fine effect is clearly beyond the precision of the present treatment for determining the abundance of PBHs. 

\section{Solving the equivalent eigenvalue problem}
\label{app:wkb}

The tail of the PDF is determined by the poles and residues of the characteristic function. As we have discussed in \Sec{sec:eigenvalues}, computing the tail of the PDF can be achieved similarly by solving an eigenvalue problem for the eigenfunctions $\Psi_n$ and eigenvalues $\Lambda_n$, that satisfy the equation
\bea
\label{eq:eigen:equation:app}
\Psi''_n-\frac{v_\phi}{v^2}\Psi'_n+\frac{\Lambda_n}{v}\Psi_n=0
\eea
with boundary conditions
\begin{align}
&\Psi_n(\phiend)=0\,, \\
&\Psi'_n(\phiuv)=0\,.
\end{align}
For generic potentials, this equation does not have an analytic solution. However, one can solve it using an adiabatic (or WKB) approximation. First, the friction term, proportional to $v_\phi$, can be absorbed through the field redefinition
\bea
\displaystyle
\Psi_n=e^{\frac{1}{2}\int\frac{v_\phi}{v^2}\dd\phi}\tilde{\Psi}_n\,.
\eea
In this way, \Eq{eq:eigen:equation:app} becomes
\bea
\tilde{\Psi}''_n+\left[\frac{\Lambda_n}{v}-\frac{1}{4}\lp\frac{v_\phi}{v^2}\rp^2+\frac{1}{2}\lp\frac{v_\phi}{v^2}\rp'\right]\tilde{\Psi}_n=0\,,
\eea
which can be solved in plane waves whenever the frequency is slowly varying. In that regime, the solution reads
\bea
\Psi_n=e^{\frac{1}{2}\int\frac{v_\phi}{v^2}\dd \phi}\lp\alpha_n e^{i\int\theta_n\dd\phi}+\beta_ne^{-i\int\theta_n\dd\phi}\rp\,,
\eea  
where the phase $\theta_n$ reads
\bea
\theta_n^2=\frac{\Lambda_n}{v}-\frac{1}{4}\lp\frac{v_\phi}{v^2}\rp^2+\frac{1}{2}\lp\frac{v_\phi}{v^2}\rp'\,.
\eea
The first boundary condition imposes
\bea
\alpha_n=-\beta_n\,,
\eea
so that the eigenfunctions are given by
\bea
\Psi_n=2i\alpha_n\, e^{\frac{1}{2}\int\frac{v_\phi}{v^2}\dd\phi}\sin\left(\int\theta_n \dd\phi\right)\,.
\eea 
The second boundary condition determines a transcendental equation for the eigenvalues $\Lambda_n$,
\bea
\tan\lb\int_{\phiend}^{\phiuv}\theta_n(\phi) \dd\phi\rb=-2\frac{v^2(\phiuv)}{v_\phi(\phiuv)}\theta_n(\phiuv)\,.
\eea
This expression provides a better approximation than the almost constant approximation applied to a vacuum-dominated potential with a constant slope in \Sec{sec:linear_potential}, see \Eq{eq:trascendental_linear}. However, one does not avoid having to solve a transcendental equation to obtain the eigenvalues.

\bibliographystyle{JHEP}
\bibliography{StochaTail}
\end{document}